\documentclass[twocolumn,english,aps]{revtex4}
\usepackage[T1]{fontenc}
\usepackage[latin9]{inputenc}
\usepackage{textcomp}
\usepackage{amsmath}
\usepackage{color}
\usepackage{graphicx}
\usepackage{amssymb}
\usepackage{esint}

\makeatletter
\@ifundefined{textcolor}{}
{%
 \definecolor{BLACK}{gray}{0}
 \definecolor{WHITE}{gray}{1}
 \definecolor{RED}{rgb}{1,0,0}
 \definecolor{GREEN}{rgb}{0,1,0}
 \definecolor{BLUE}{rgb}{0,0,1}
 \definecolor{CYAN}{cmyk}{1,0,0,0}
 \definecolor{MAGENTA}{cmyk}{0,1,0,0}
 \definecolor{YELLOW}{cmyk}{0,0,1,0}
 }

\makeatother

\makeatother

\usepackage{babel}

\begin{document}

\preprint{This line only printed with preprint option}

\title{Time Constants of Spin-Dependent Recombination Processes}

\author{Felix Hoehne}
\email[corresponding author, email: ]{hoehne@wsi.tum.de}

\affiliation{Walter Schottky Institut, Technische Universit\"{a}t M\"{u}nchen, Am Coulombwall
4, 85748 Garching, Germany}

\author{Lukas Dreher}

\affiliation{Walter Schottky Institut, Technische Universit\"{a}t M\"{u}nchen, Am Coulombwall
4, 85748 Garching, Germany}

\author{Max Suckert}

\affiliation{Walter Schottky Institut, Technische Universit\"{a}t M\"{u}nchen, Am Coulombwall
4, 85748 Garching, Germany}

\author{David P. Franke}

\affiliation{Walter Schottky Institut, Technische Universit\"{a}t M\"{u}nchen, Am Coulombwall
4, 85748 Garching, Germany}

%
%
%

\author{Martin Stutzmann}

\affiliation{Walter Schottky Institut, Technische Universit\"{a}t M\"{u}nchen, Am Coulombwall
4, 85748 Garching, Germany}

\author{Martin S.~Brandt}

\affiliation{Walter Schottky Institut, Technische Universit\"{a}t M\"{u}nchen, Am Coulombwall
4, 85748 Garching, Germany}
\begin{abstract}
We present experiments to systematically study the time constants of spin-dependent recombination processes in semiconductors using pulsed electrically detected magnetic resonance (EDMR). The combination of time-programmed optical excitation and pulsed spin manipulation allows us to directly measure the recombination time constants of electrons via localized spin pairs and the time constant of spin pair formation as a function of the optical excitation intensity. Using electron nuclear double resonance, we show that the time constant of spin pair formation is determined by an electron capture process. Based on these time constants we devise a set of rate equations to calculate the current transient after a resonant microwave pulse and compare the results with experimental data. Finally, we critically discuss the effects of different boxcar integration time intervals typically used to analyze pulsed EDMR experiments on the determination of the time constants. The experiments are performed on phosphorus-doped silicon, where EDMR via spin pairs formed by phosphorus donors and Si/SiO$_2$ interface dangling bond defects is detected.
\end{abstract}
\maketitle

\section{introduction}
Electrically detected magnetic resonance (EDMR) has been widely used as a high-sensitivity alternative to conventional electron spin resonance (ESR) to study paramagnetic point defects in semiconductors~\cite{Lepine72Spindep, Stutzmann2000,Schnegg2012}. A large class of spin-dependent recombination processes investigated with EDMR is based on the formation and recombination of weakly coupled spin pairs~\cite{Kaplan78Spindep,Stich1995,Spaeth03}. A prototype example of such a spin pair recombination process is observed in phosphorus-doped crystalline Si, where paramagnetic interface defects (e.g. P$_\text{b}$-centers) at the Si/SiO$_2$ interface and $^{31}$P donors in their vicinity constitute the spin pair (Fig.~\ref{fig:RateScheme})~\cite{Stegner06,Hoehne10}. Detailed understanding and modeling of such recombination mechanisms require the knowledge of the time constants involved in the different steps of the recombination process~\cite{Movaghar1980, Barabanov1996}. This is especially important for
the design of complex pulse sequences like, e.g., pulsed electrically detected electron nuclear double resonance (ENDOR)~\cite{HoehneENDOR2011,Dreher2012}. Using continuous wave (cw) EDMR measurements, these time constants can be inferred, if at all, only indirectly via, e.g., a variation of the magnetic field modulation frequency or the microwave field amplitude~\cite{Dersch83}. Here, pulsed EDMR under cw illumination~\cite{Boehme03EDMR, Stegner06} offers a more direct way of accessing some of these time constants, e.g., by measuring the current transient after an excitation of the spin system by a short microwave pulse~\cite{Boehme03EDMR} or by applying special pulse sequences~\cite{Paik2010T1T2, Behrends09}. However, these approaches suffer from several drawbacks like the difficulty of separating the different time constants involved and the influence of the bandwidth of the detection setup on the observed time constants.   

In this paper, we use pulsed EDMR in combination with time-programmed optical excitation to determine the time constants of the spin-dependent recombination process via $^{31}$P-P$_\text{b0}$ spin pairs. This approach not only allows us to devise experiments which access the time constants of spin pair recombination and formation separately. In addition to allowing to measure the recombination rate of antiparallel spin pairs, this approach also makes measurements of the recombination rate of parallel triplet spin pairs possible, which have not been reported yet, and allows for the discrimination between electron and hole capture processes for the spin pair formation via pulsed ENDOR. This paper significantly extends the measurements of the spin pair recombination rates as presented in Ref.~\cite{Dreher2012}.

The paper is organized as follows: In Sect.~\ref{ss:recombination process}, we first discuss the recombination process via weakly coupled spin pairs in general. We devise a set of rate equations to calculate the temporal evolution of the spin system in Sect.~\ref{ss:rateequations}. In Sect.~\ref{ss:ExpDetails}, we describe the sample and the measurement setup and discuss in some detail the spin-to-charge conversion for pulsed EDMR with time-programmed illumination. In the main part of the paper, we describe experiments to measure the recombination rate of antiparallel spin pairs (Sect.~\ref{ss:rs}), the
generation rate of new spin pairs (Sect.~\ref{ss:generation}), and the recombination rate of parallel spin
pairs (Sect.~\ref{ss:rt}). We then use the time constants determined in Sect.~\ref{results} in combination with a rate equation model to calculate the current transient after a resonant microwave pulse and compare the results with experimental data (Sect.~\ref{ss:transients}). Finally, we discuss the implications of the variation of recombination time constants over the ensemble of spin pairs under investigation (Sect.~\ref{ss:distribution}).

\section{Recombination Process}
\label{ss:recombination process}

\begin{figure}[ht!]
\begin{centering}
\includegraphics[width=8cm]{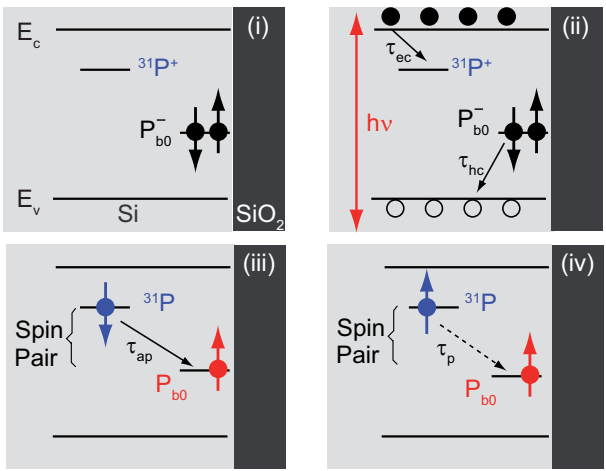}
\par\end{centering}
\caption{\label{fig:RateScheme}
Illustration of the recombination process via $^{31}$P-P$_\mathrm{b0}$ spin pairs. (i) Without illumination the phosphorus donors are in their positive charge state $^{31}$P$^+$ because of the compensation by the negatively charged P$_\mathrm{b0}^-$. 
(ii) Upon illumination electrons (holes) are captured by the $^{31}$P$^+$ (P$_\mathrm{b0}^{-}$) with a rate $1/\tau_\mathrm{ec}$ ($1/\tau_\mathrm{hc}$) resulting in the generation of new spin pairs with a rate $1/\tau_\mathrm{g}=1/\tau_\mathrm{ec}+1/\tau_\mathrm{hc}$. For photo-neutralized phosphorus donors and P$_\mathrm{b0}$ centers in spatial proximity a transition of the donor electron to the P$_\mathrm{b0}$ can take place with a fast rate 1/$\tau_\mathrm{ap}$ for antiparallel spin pairs (iii) and a slow rate 1/$\tau_\mathrm{p}$ for parallel spin pairs (iv). 
} 
\end{figure}

In the following, we summarize the basic features of the spin pair recombination process exemplarily for the $^{31}$P-P$_\mathrm{b0}$ spin pair, defining the relevant time constants
as depicted in Fig.~\ref{fig:RateScheme}. This picture is based on the model of weakly coupled spin pairs proposed by Kaplan, Solomon and Mott (KSM model)~\cite{Kaplan78Spindep} and elaborated in Ref.~\cite{White1977,Movaghar1980,Haberkorn1980,Lips94,Barabanov1996}.
It has been established that this $^{31}$P-P$_\mathrm{b0}$ process is the dominant spin-dependent recombination process for phosphorus donors near the Si/SiO$_2$ interface~\cite{Hoehne10}.
We assume that, without illumination, the $^{31}$P
donors at the Si/SiO$_2$ interface are compensated by interface
defects and therefore are in the ionized $^{31}$P$^+$ state as
sketched in panel (i). Upon illumination (ii), electrons are captured by the $^{31}$P$^+$ donors with a time constant $\tau_\mathrm{ec}$ and holes by the P$_\mathrm{b0}^-$ with a time constant $\tau_\mathrm{hc}$   
forming $^{31}$P-P$_\mathrm{b0}$ spin pairs [(iii) and (iv)] with a generation rate $1/\tau_\mathrm{g}=1/\tau_\mathrm{ec}+1/\tau_\mathrm{hc}$.
The spin pair will return to the $^{31}$P$^+$-P$_\mathrm{b0}^-$ state (i) on a time
scale of $\tau_\mathrm{ap}$ for antiparallel spin configuration (iii) or will remain stable on a much longer time scale $\tau_\mathrm{p}$ for parallel
spin orientation (iv) because of the Pauli principle. Consequently, a dynamic equilibrium
is established, in which in good approximation all of the
spin pairs are in the parallel configuration, which we refer
to as the steady state. As will be shown in this work, $\tau_\mathrm{p}/\tau_\mathrm{ap}\approx$100 in the samples studied here, so that only a fraction of $\approx$0.01 of the spin pairs is in an antiparallel state. ESR-induced transitions of either the $^{31}$P or the P$_\mathrm{b0}$ electron spin lead to a net transformation of parallel into antiparallel spin pairs and therefore increase the overall recombination rate, giving rise to the resonant photocurrent quenching observed in EDMR.

In the picture of the recombination process presented above, we neglect the possibility of spin pair dissociation through excitation of an electron into the conduction band, a key feature of the KSM model~\cite{Kaplan78Spindep}, since at low temperatures ($T$=5~K) thermal excitation can be neglected and impact ionization is not expected to play a role during time intervals without optical excitation due to the lack of carriers in the conduction and valence bands. We also neglect the effects of spin decoherence, since typical decoherence times in the $^{31}$P-P$_\mathrm{b0}$ spin system are of the order of several~$\mu$s~\cite{Huebl08Echo,Paik2010T1T2}, long compared with the timescale of the detection echoes used below and shorter than the other time constants of the recombination process. This allows us to significantly simplify the discussion of the spin pair dynamics in Sect.~\ref{ss:rateequations} when compared with Ref.~\cite{Boehme02PhD,Huebl07PhD} by considering only the populations of the different states of the spin system.  

\section{Rate Equation Model}
\label{ss:rateequations}

In this section we describe a set of rate equations modeling the dynamics of the spin-dependent recombination process. The spin system under consideration consists of the phosphorus electron spin ($S$=1/2), the phosphorus nuclear spin ($I$=1/2) and the dangling bond electron spin ($S$=1/2), so that in general 8 basis states are needed to describe its dynamics. The discussion can be simplified by first neglecting the nuclear spin degree of freedom, since the recombination dynamics are governed by the electron spin states and the electron Zeeman interaction is much larger than the hyperfine interaction in the experiments presented here, so that state mixing can be neglected. The number of basis states can be further reduced by considering only the relative orientation of the $^{31}$P and P$_\mathrm{b0}$ spins, neglecting their orientation with respect to the external magnetic field $B_0$. This is possible since the recombination dynamics of the spin pairs only depend on the mutual orientation of the two spins forming the spin pair~\cite{Kaplan78Spindep}. This manifests itself, e.g., in the magnetic field-independence of the EDMR signal amplitude~\cite{Solomon1977,Brandt98Linewidth, Baker2012}. These simplifications reduce the number of involved states to two, with the fraction of parallel and antiparallel spin pairs denoted by $n_\mathrm{p}$ and $n_\mathrm{ap}$, respectively. An additional state $n_+$ is introduced to quantify the fraction of spin pairs with ionized $^{31}$P$^+$ donors and doubly occupied P$_\mathrm{b0}^-$. We assume that the spin pair is always in one of these three states giving rise to the normalization condition $n_\mathrm{p}+n_\mathrm{ap}+n_\mathrm{+}$=1.

Based on the recombination picture shown in Fig.~\ref{fig:RateScheme} we can establish a system of rate equations given by
\begin{equation}
\label{eq:rateequations}
\frac{\text{d}}{\text{d}t}\rho=\tilde{R}\rho,
\end{equation}
with
\begin{equation}
\label{eq:rho}
\rho=
\begin{pmatrix}
n_\mathrm{p} \\
n_\mathrm{ap} \\
n_\mathrm{+} \\
\end{pmatrix},
\end{equation}
and
\begin{equation}
\label{eq:recombmatrix}
\tilde{R}=
\begin{pmatrix}
-1/\tau_\mathrm{p} & 0 & 1/2\tau_\mathrm{g}\\
0 & -1/\tau_\mathrm{ap} & 1/2\tau_\mathrm{g}\\
1/\tau_\mathrm{p} & 1/\tau_\mathrm{ap} & -1/\tau_\mathrm{g}
\end{pmatrix}.
\end{equation}  
The typical coupling of \textless 1~MHz of the spin pairs studied here~\cite{Suckert2013} is small compared to the Zeeman energy, so that the eigenstates of the $^{31}$P-P$\mathrm{b0}$ spin pair are in good approximation the product states. We also assume that all eigenstates are generated with equal probability, so that the generation rate for the parallel and antiparallel state is $1/2\tau_\text{g}$, in contrast to the probability of three to one of forming a triplet or a singlet state for strongly coupled spins as observed, e.g., in organic semiconductors~\cite{Wilson2001,Reufer2005}. In Sect.~\ref{ss:transients}, we further elaborate this model by dividing the generation of new spin pairs into separate electron and hole capture processes. 

The steady state solution $\frac{\text{d}}{\text{d}t}\rho_\mathrm{eq}$=0 of Eq.~\eqref{eq:rateequations} is given by
\begin{equation}
\rho_\mathrm{eq}=
\begin{pmatrix}
n^\mathrm{eq}_\mathrm{p} \\
n^\mathrm{eq}_\mathrm{ap} \\
n^\mathrm{eq}_\mathrm{+} \\
\end{pmatrix}
= \frac{1}{1+\frac{1}{2\tau_\mathrm{g}}(\tau_\mathrm{p}+\tau_\mathrm{ap})}
\begin{pmatrix}
 \frac{\tau_\mathrm{p}}{2\tau_\mathrm{g}}\\
 \frac{\tau_\mathrm{ap}}{2\tau_\mathrm{g}} \\
1 \\
\end{pmatrix}.
\end{equation} 
The time evolution of the spin pair ensemble can be calculated by 
\begin{equation}
\rho(t) = e^{\tilde{R}t}\cdot\rho(0),
\end{equation}
where $\rho(0)$ denotes the initial state of the system.
The characteristic time constants of the temporal evolution of $\rho(t)$ are determined by the inverse eigenvalues $\lambda_\mathrm{i}$ of the matrix $\tilde{R}$. These eigenvalues are given by
\begin{equation}
\label{eq:time constants_eigenvalues}
\begin{split}
\lambda_1 &= -\frac{1}{2}\left(\frac{1}{\tau_\mathrm{g}}+\frac{1}{\tau_\mathrm{ap}}+\frac{1}{\tau_\mathrm{p}}-\sqrt{\left(\frac{1}{\tau_\mathrm{g}}\right)^2+\left(\frac{1}{\tau_\mathrm{ap}}-\frac{1}{\tau_\mathrm{p}}\right)^2}\right) \\
\lambda_2 &= -\frac{1}{2}\left(\frac{1}{\tau_\mathrm{g}}+\frac{1}{\tau_\mathrm{ap}}+\frac{1}{\tau_\mathrm{p}}+\sqrt{\left(\frac{1}{\tau_\mathrm{g}}\right)^2+\left(\frac{1}{\tau_\mathrm{ap}}-\frac{1}{\tau_\mathrm{p}}\right)^2}\right) \\
\lambda_3 &= 0.
\end{split}
\end{equation}
If the recombination of antiparallel spin pairs is much faster than the recombination of parallel spin pairs and the generation of new spin pairs, the expressions~\eqref{eq:time constants_eigenvalues} simplify to
\begin{equation}
\label{eq:time constants_eigenvalues_simple}
\begin{split}
\lambda_1 &= -\frac{1}{2}\left(\frac{1}{\tau_\mathrm{g}}+\frac{2}{\tau_\mathrm{p}}\right) \\
\lambda_2 &= -\frac{1}{\tau_\mathrm{ap}} \\
\lambda_3 &= 0.
\end{split}
\end{equation}
In the case that the optical excitation is switched off, no new spin pairs are generated, so that $\frac{1}{\tau_\mathrm{g}}=0$, which simplifies Eqs.~\eqref{eq:time constants_eigenvalues} further to
\begin{equation}
\label{eq:time constants_eigenvalues_simple_simple}
\begin{split}
\lambda_1 &= -\frac{1}{\tau_\mathrm{p}} \\
\lambda_2 &= -\frac{1}{\tau_\mathrm{ap}} \\
\lambda_3 &= 0.
\end{split}
\end{equation}
Under these conditions the characteristic time constants of the temporal evolution of the spin system are given by the recombination times of parallel and antiparallel spin pairs, $\tau_\mathrm{p}$ and $\tau_\mathrm{ap}$, respectively. This is the reason why pulsed optical excitation is advantageous for the characterization of the recombination time constants.

If the simplifications discussed above cannot be made, e.g., for the inversion recovery experiment under continuous optical excitation discussed in Sect.~\ref{ss:rs}, the temporal evolution of the spin system can still be calculated analytically.
The resulting expressions are, however, lengthy and do not provide additional insight.
We therefore will resort to a numerical solution of Eq.~\eqref{eq:rateequations} to describe experiments for which a straightforward assignment of time constants is difficult.

\section{Experimental Details}
\label{ss:ExpDetails}

\subsection{Spin-to-Charge Conversion}
\label{ss:pulsed readout}
In all experiments presented in this work, we use the amplitude of a spin echo~\cite{Huebl08Echo} to measure the spin state of the spin pair. In particular, we employ a two-step phase cycling sequence where the phase of the last $\pi/2$ projection pulse is switched by 180$^{\circ}$ denoted as ($\pm x$)~\cite{Hoehne2012}. Depending on the phase of the projection
pulse, the detection echo sequence forms an effective 2$\pi$ pulse for (+x) or an effective $\pi$ pulse for (-x) and the signal amplitude is obtained by taking the difference of the signals obtained in the two cycles as discussed below. For spin systems with a sufficient phase memory time, the spin echo can be replaced by $\pi$/2-$\pi$/2 (Ramsey) pulse sequence for which the same phase cycling sequence can be employed~\cite{Lu2011}. The phase cycling sequence removes background signals resulting from spin-independent current transients caused by the microwave and light pulses. In addition, it is used to realize a lock-in detection scheme by switching the phase at frequencies between $1-100$\,Hz resulting in a tenfold improvement of the signal-to-noise ratio~\cite{Hoehne2012}. However, small differences in the amplitude of the microwave pulses with (+x) and (-x) phase result in an incomplete background removal. This can be mitigated by extending the phase cycling sequence to a 4-step phase cycle shown in Table~\ref{table1} for which also the phase of the first microwave $\pi$/2 pulse is switched between (+x) and (-x). The additional cycles 3 and 4 constitute the same effective pulses as the cycles 2 and 1, as can be seen by the corresponding effective pulse lengths. However, we limit the following discussion by considering only the cycles 1 and 2.  

\begin{figure}[ht!]
\begin{centering}
\includegraphics[width=8cm]{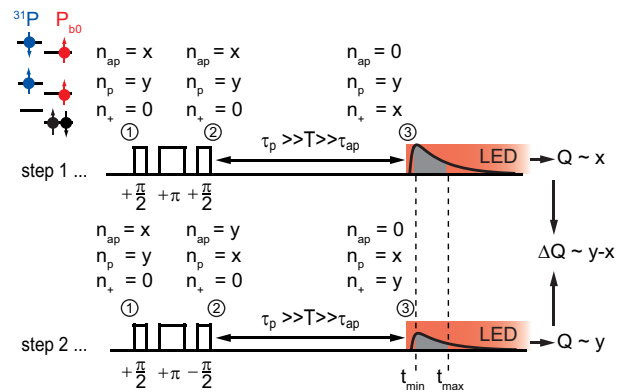}
\par\end{centering}
\caption{\label{fig:PulsedReadout} 
Readout spin echo pulse sequence for pulsed optical excitation. Boxcar integration (gray shaded area) of the photocurrent transient after switching on the illumination and subtracting step 2 from step 1 results in a charge $\Delta Q$ proportional to the difference between the number of antiparallel
and parallel spin pairs at the beginning of the readout pulse sequence. The number of spin pairs in the antiparallel, parallel,
and the $^{31}$P$^+$-P$_\text{b0}$ state is denoted by $n_\text{ap}$, $n_\text{p}$ and $n_\text{+}$, respectively.}

\end{figure}

To describe the spin-to-charge conversion under time-programmed optical excitation~\cite{Dreher2012},
we discuss the dynamics of the $^{31}$P-P$_\text{b0}$ electron-spin pair in terms of the three states depicted in Fig.~\ref{fig:RateScheme}
(i), (iii), and (iv). 
Assuming that at the
beginning of the pulse sequence at time (1) (Fig.~\ref{fig:PulsedReadout}) there are $n_\text{ap}$=$x$ antiparallel spin pairs and $n_\text{p}$=$y$ parallel spin pairs,
a spin echo forming an effective 2$\pi$ pulse leaves the states unaffected. During the time interval $T$, chosen
such that $\tau_\text{p}\gg T\gg\tau_\text{ap}$, all antiparallel spin pairs are transferred into the $^{31}$P$^+$-P$_\text{b0}^-$ state, while the parallel spin
pairs essentially remain unchanged, resulting in $n_\text{ap}$=0, $n_\text{p}$=$y$ and $n_\text{+}$=$x$ at time (3). After switching on the light,
\begin{table}%
\caption{Detection spin echo phase cycling sequence used to readout the spin system. Here, +x and -x denote the phases of the microwave pulses which are shifted by 180$^\circ$ with respect to each other.}
\label{table1}
\begin{tabular}{c|c|c|c|c}
\hline \hline
step & $\pi$/2 & $\pi$ & $\pi$/2 & effective pulse length\\
\hline
1 & +x & +x  & +x & 2$\pi$  \\
2 & +x & +x  & -x & $\pi$  \\
3 & -x & +x  & +x & $\pi$  \\
4 & -x & +x  & -x & 0  \\
\hline \hline
\end{tabular}
\end{table}
a current transient occurs. Its spin-dependent part reflects the generation of new $^{31}$P-P$_\text{b0}$ spin pairs
and the spin-dependent amplitude is therefore proportional to $n_\text{+}$=$x$, the number of antiparallel spin pairs at time
(1) before the detection pulse sequence. Repeating the same pulse sequence with a spin echo forming an effective $\pi$
pulse results in a current transient with its spin-dependent amplitude proportional to $n_\text{+}$=$y$, the number of parallel
spin pairs at time (1), as shown in Fig.~\ref{fig:PulsedReadout}. A large portion of the photocurrent transient, induced by the onset of
the LED illumination, is spin-independent and thus is independent of the phases of the applied microwave pulses; it is removed
when step 2 is subtracted from step 1. Boxcar integration of the subtracted photocurrent transients from $t_\mathrm{min}$ to $t_\mathrm{max}$ results in a charge $\Delta Q$, which is
proportional to the difference between the number of antiparallel and parallel spin pairs before the echo sequence. A similar reasoning can be applied to the case of continuous illumination, so that again $\Delta Q\propto n_\text{ap}-n_\text{p}$ before the echo sequence~\cite{Boehme03EDMR}.
%


\subsection{Sample and Setup}
\label{ss:SS}

The sample used in this work is fabricated by chemical vapor deposition
and consists of a nominally $22$\,nm-thick (001)-oriented Si layer with a natural isotope composition and
$\text{[P]}=3\cdot10^{16}$\,cm$^{-3}$ covered by a native
oxide. It is grown on a $2.5$\,\textmu{}m thick, nominally undoped
$^{\text{nat}}$Si buffer on a silicon-on-insulator (SOI) substrate. Evaporated
interdigit Cr/Au contacts with a period of $20$\,\textmu{}m
covering an active area of $2\times2.25$\,mm$^{2}$ are biased with
$300$\,mV. The sample is illuminated with a red LED (630~nm) with an illumination intensity of 20~mW/cm$^2$ if not stated otherwise, measured by a photodetector placed at the sample position inside the resonator. Pulsed illumination is achieved by modulating the LED current using a Thorlabs LDC 210C current controller providing pulse rise and fall times of typically 2~$\mu$s.
The measurements are performed
at $5.0$~K stabilized to $\pm$0.1~K in a helium gas flow cryostat. The samples are oriented in an external magnetic field $B_0$ with the [110] axis of the Si wafer parallel to $B_0$. 

The pulsed EDMR experiments are performed at a microwave frequency
of $\nu_{\text{mw}}=9.739$\,GHz in a Bruker X-band dielectric microwave resonator for pulsed ENDOR at $B_0$=350.65~mT chosen such that the microwave pulses resonantly excite the high-field line of the hyperfine split $^{31}$P electron spin transitions. The corresponding EDMR spectrum can be found, e.g., in Ref.~\cite{Lu2011}. The
microwave and rf pulses are shaped using a SPINCORE PulseBlasterESR-Pro 400
MHz pulse generator and a system of microwave mixers. The mw pulses are
amplified by an Applied Systems Engineering 117X traveling wave tube
with a maximum peak power of 1\,kW. The microwave power is adjusted to achieve
a $\pi$ pulse time of $\tau_{\pi}=40$\,ns ($B_{1}=0.45$\,mT) for the spin echoes and $\tau_{\pi}=30$\,ns ($B_{1}=0.6$\,mT) for the inversion pulses to ensure that the bandwidth of the inversion pulse is larger than the detection bandwidth~\cite{Schweiger01}. The length of the free evolution intervals of the spin echoes is $\tau_1=\tau_2$=300~ns. The ENDOR rf pulses are amplified by a 300~W solid state amplifier resulting in a $\pi$ pulse time of 29~$\mu$s for the $^{31}$P$^+$ nuclear spin transition at 6.036~MHz. The current transients after the pulse sequence are amplified by a custom-built balanced transimpedance amplifier (Elektronik-Manufaktur, Mahlsdorf) with low- and high-pass filtering at cut-off frequencies of 2~kHz and 1~MHz, respectively, and recorded with a fast data acquisition card (Gage). The cuurent transients are box-car integrated over an interval from typically 5~$\mu$s to 40~$\mu$s resulting in a charge $Q$. The measurements are repeated and averaged with different shot repetition times chosen such that the sample is illuminated for at least 5~ms before each pulse sequence to ensure that at the beginning of the pulse sequence the spin system has reached a steady state.

\section{Measurement of the Recombination Time constants}
\label{results}
In the following main part of this paper, we present different pulse sequences combining pulsed illumination and spin excitation to measure the recombination rates of parallel and antiparallel spin pairs, as well as the generation rate of new spin pairs.

\subsection{Recombination Rate of Antiparallel Spin Pairs}
\label{ss:rs}

\begin{figure}[t!]
 \begin{centering}
\includegraphics[width=8cm]{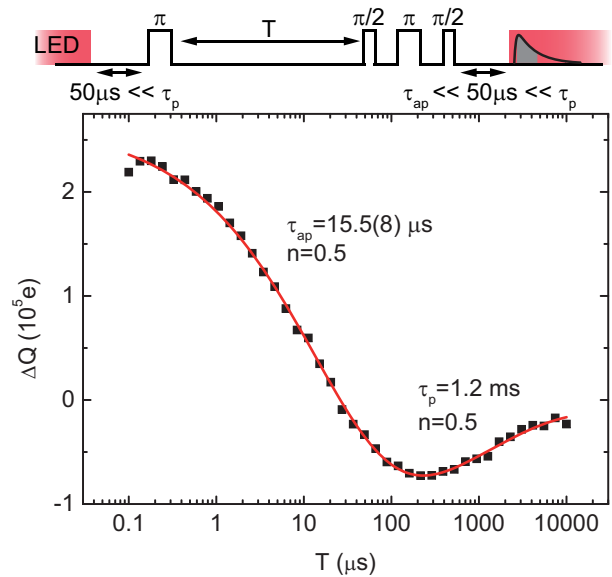} 
\par\end{centering}
\caption{\label{fig:InvRecDark}
In the upper part, the pulse sequence used to determine the recombination time of antiparallel spin pairs $\tau_\mathrm{ap}$ is depicted. It is based on a standard inversion recovery pulse sequence with spin echo detection. To separate the effects of spin pair recombination and generation, the above bandgap illumination provided by a red LED is switched off 50~$\mu$s before the inversion pulse and switched on 50~$\mu$s after the detection spin echo, so that during the inversion recovery pulse sequence no new spin pairs are generated. 
In the lower part, the amplitude of the phase-cycled detection echo $\Delta Q$ (black symbols) is shown as a function of the waiting time $T$ between the inversion pulse and the detection spin echo. The observed decay can be fitted with two stretched exponential decays with a constant offset (red line), the first with a time constant of $\tau_\mathrm{ap}$=15.5(8)~$\mu$s and an exponent of $n$=0.5. The second decay with a longer time constant of $\tau_\mathrm{p}$=1.2~ms and $n$=0.5 reflects the recombination of those spin pairs for which the $^{31}$P spin has not been inverted by the first $\pi$ pulse and which therefore are in a parallel state during the waiting time $T$. The small offset is caused by imperfections in the background subtraction of the phase cycling sequence. 
}
\end{figure}

Following the discussion in Sect.~\ref{ss:rateequations}, we have seen that it is advantageous to switch off the optical excitation to separate the effects of spin pair recombination and generation. In this case, the characteristic time constants of the temporal evolution of the spin system are given by $\tau_\mathrm{ap}$ and $\tau_\mathrm{p}$ [see Eq.~\eqref{eq:time constants_eigenvalues_simple_simple}]. To measure the recombination time of antiparallel spin pairs~$\tau_\mathrm{ap}$, we therefore employ a standard inversion recovery pulse sequence~\cite{Schweiger01} and switch off the optical excitation during the pulse sequence as sketched in Fig.~\ref{fig:InvRecDark}.
Under illumination almost all spin pairs are in the parallel state and remain so for a time $\tau_\mathrm{p}$ after switching off the LED. The inversion $\pi$ pulse is applied 50~$\mu$s after switching off the illumination which is much longer than the 2~$\mu$s fall time of the light pulse and the carrier lifetime (cf. Sect.~\ref{ss:generation}) to ensure that no carriers are left in the conduction band. The $\pi$ pulse creates antiparallel spin pairs which recombine during the waiting time $T$, however, in contrast to the case of continuous illumination, no new spin pairs are generated. Therefore, $\tau_\mathrm{ap}$ can be determined directly by measuring the number of antiparallel spin pairs as a function of $T$ using the spin echo sequence followed by a detection light pulse as discussed in Sect.~\ref{ss:pulsed readout}.

The result of such a measurement is shown in Fig.~\ref{fig:InvRecDark}, where the detection echo amplitude $\Delta Q\propto n_\mathrm{ap}-n_\mathrm{p}$ is plotted as a function of $T$ (black squares). The first decay reflects the recombination of antiparallel spin pairs. It can be fitted with a stretched exponential $\Delta Q\propto e^{-(T/\tau_\mathrm{ap})^n}$ with a time constant of $\tau_\mathrm{ap}$=15.5(8)~$\mu$s and an exponent of $n$=0.5 (red line). The stretched exponential character of the decay is thought to be the consequence of the broad distribution of recombination rates within the ensemble of spin pairs caused by a distribution of $^{31}$P-P$_\text{b0}$ distances within the sample. An overshoot is observed followed by a second decay with a longer time constant of $\tau_\mathrm{p}$=1.2~ms and $n$=0.5, reflecting the recombination of those spin pairs for which the $^{31}$P spin has not been inverted by the first $\pi$ pulse and which are therefore in a parallel state. They recombine with a rate $1/\tau_\mathrm{p}$ during the waiting time $T$. The bandwidth of the $\pi$ pulse ($B_1$=0.6~mT) is not sufficient to fully excite the inhomogeneously broadened $^{31}$P electron spin transition with a peak-to-peak linewidth of 0.3~mT~\cite{Lu2011}. To reduce uncertainties in the determination of these rates, we use further experiments discussed in Sect.~\ref{ss:rt} to independently measure $\tau_\mathrm{p}$. The time constant and exponent of the slower decay of the fit were therefore fixed to the values obtained in those experiments and only $\tau_\mathrm{ap}$ and the amplitudes of the decays were left as free fitting parameters. 

The stretched exponential character of the decay indicates that recombination processes with a broad distribution of time constants rather than a single time constant are observed. We attribute this to a variation of the $^{31}$P-P$_\mathrm{b0}$ distance over the spin pair ensemble under investigation. Since the spin-dependent transition between the localized donor and defect states involves a tunneling process, even small variations of the tunneling distance result in a broad distribution of recombination time constants. Measurements of the exchange coupling between the $^{31}$P donor and the P$_\mathrm{b0}$ using electrically detected double electron electron resonance (DEER) reveal coupling strengths around 600~kHz, compatible with a distribution of spin pair distances ranging from 14~nm to 20~nm~\cite{Suckert2013}.

\begin{figure}[t!]
\begin{centering}
\includegraphics[width=8cm]{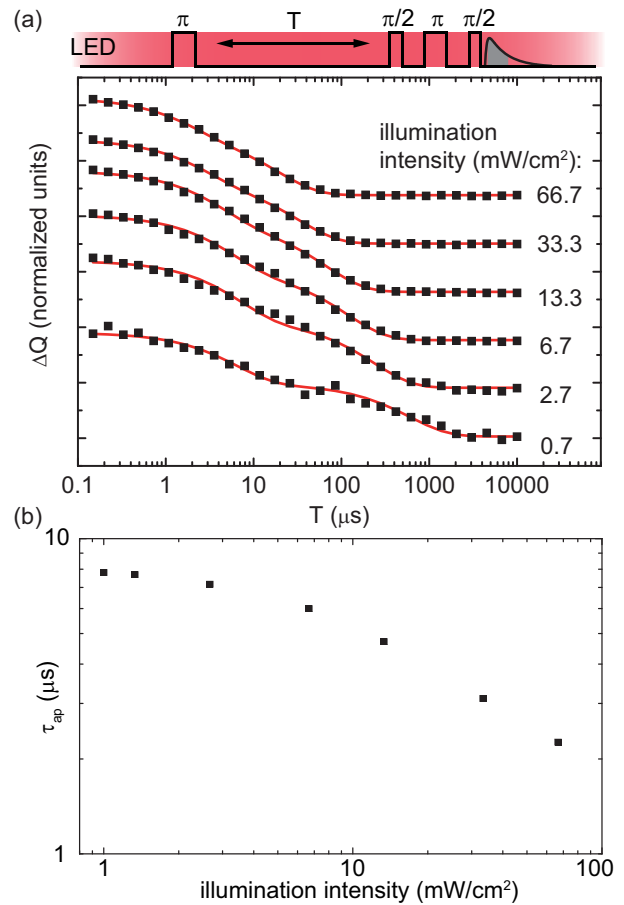} 
\par\end{centering}
\centering{}\caption{\label{fig:InvRecIllu}
Inversion recovery pulse sequence under continuous illumination. After a mw $\pi$ pulse, the spin pairs are dominantly in an antiparallel state from where they recombine with a time constant $\tau_\mathrm{ap}$ during the waiting time $T$, while at the same time, new spin pairs are generated with a rate $1/\tau_\mathrm{g}$. 
The phase-cycled echo amplitudes are shown in panel (a) for different light intensities (black symbols, data offset for clarity). $\Delta Q$ is given in normalized units to allow a direct comparison of the data traces for different illumination intensities. The data traces are fitted by numerically solving the system of rate equations (\ref{ss:rateequations}) with $\tau_\mathrm{g}$ and $\tau_\mathrm{ap}$ as fitting parameters. The resulting fits are shown as red lines with the different recombination times $\tau_\text{ap}$ plotted in (b). The generation rates $1/\tau_\mathrm{g}$ are summarized in Fig.~\ref{fig:g_summary}.}
\end{figure}

It is interesting to compare these results with a standard inversion recovery experiment under continuous illumination which has been used previously to determine the recombination rate of antiparallel spin pairs~\cite{Paik2010T1T2}. The corresponding pulse sequence is depicted in Fig.~\ref{fig:InvRecIllu}(a).
In contrast to the case of pulsed illumination, here, new spin pairs are formed during the waiting time $T$ by electron and hole capture processes with a time constant $\tau_\mathrm{g}$, resulting in a recovery of the echo amplitude for long $T$.

The results of such a measurement are shown in Fig.~\ref{fig:InvRecIllu}(a) for different illumination intensities stated next to the traces.
 All traces have been normalized and offset to allow for easier comparison of the involved time constants. Due to the interplay of both $\tau_\mathrm{ap}$ and $\tau_\mathrm{g}$, the observed decay cannot be described by a single stretched exponential dependence. However, already by comparing, e.g., the top and bottom traces it can be seen that the observed time constants strongly depend on the illumination intensity. Moreover, for the lowest values of the illumination intensity [bottom traces in Fig.~\ref{fig:InvRecIllu}(a)], a step in the $\Delta Q$ trace is clearly observed, indicating the presence of two different time constants.
 
We attribute the shorter time constant to the recombination of antiparallel spin pairs and the longer time constant to the generation of new spin pairs. This assignment is motivated by the fact that a recombination process has to take place before a new spin pair can be generated. However, for most of the traces a straightforward discrimination of the two time constants is not possible and we have to resort to modeling the decay with the system of rate equations introduced in Sect.~\ref{ss:rateequations}.

To this end,
we start with the state vector $\rho(0)$ after the inversion $\pi$ pulse given by
\begin{equation}
\rho(0) = 
\begin{pmatrix}
0 & 1 & 0\\
1 & 0 & 0\\
0 & 0 & 1\\
\end{pmatrix}
\rho_\mathrm{eq},
\end{equation}   
and further calculate
\begin{equation}
\label{eq:time constants_InvRec}
\rho(T) = e^{\tilde{R}\cdot T}\rho(0),
\end{equation} 
where $T$ denotes the time interval between the inversion pulse and the detection echo and $\rho(T)$ denotes the state vector before the detection echo. The amplitude of the detection echo is given by $n_\mathrm{ap}(T)-n_\mathrm{p}(T)$ as discussed in Sect.~\ref{ss:pulsed readout}.

We fit the data shown in Fig.~\ref{fig:InvRecIllu}(a) with a numerical solution of Eq.~\eqref{eq:time constants_InvRec} (red lines) with fitting parameters $\tau_\mathrm{ap}$, $\tau_\mathrm{g}$, and an amplitude and an offset for each value of the illumination intensity. The offset accounts for an imperfect subtraction of the background by the lock-in detection scheme. In addition, we use fixed values of $\tau_\mathrm{p}$=1.2~ms obtained from the experiments described in Sections~\ref{ss:rs} and~\ref{ss:rt}. The fits reproduce the basic features of the data traces quite well, although the step-like structures appear more pronounced in the fits, e.g. for 2.7~mW/cm$^{2}$ and 6.7~mW/cm$^{2}$ in Fig.~\ref{fig:InvRecIllu}(a). This is attributed to the fact that the solution of Eq.~\eqref{eq:time constants_InvRec} involves only standard exponential decays thereby ignoring the stretched exponential character observed in Fig.~\ref{fig:InvRecDark}, which masks the step-like structure.

The values of $\tau_\text{ap}$ extracted from the fits are plotted in Fig.~\ref{fig:InvRecIllu}(b). We find a value of $\tau_\text{ap}\approx$7~$\mu$s for low illumination intensities which slightly decreases for higher illumination intensities, possibly due to different subensembles of spin pairs contributing to the observed signal at higher illumination intensities. The values of $\tau_\text{ap}$ observed under illumination agree within a factor of two with the values reported in Ref.~\cite{Paik2010T1T2} for the $^{31}$P-P$_\mathrm{b0}$ spin pair. The difference between $\tau_\text{ap}$ under illumination and in the dark is again mainly related to the stretched exponential character of the decay.  
The values for $1/\tau_\mathrm{g}$ obtained here are summarized in Fig.~\ref{fig:g_summary} and discussed in section~\ref{ss:generation} together with the result from further experiments performed to measure $\tau_\mathrm{g}$.

%
%
%

\subsection{Generation Rate of Spin Pairs}
\label{ss:generation}

\begin{figure}[t!]
\begin{centering}
\includegraphics[width=8cm]{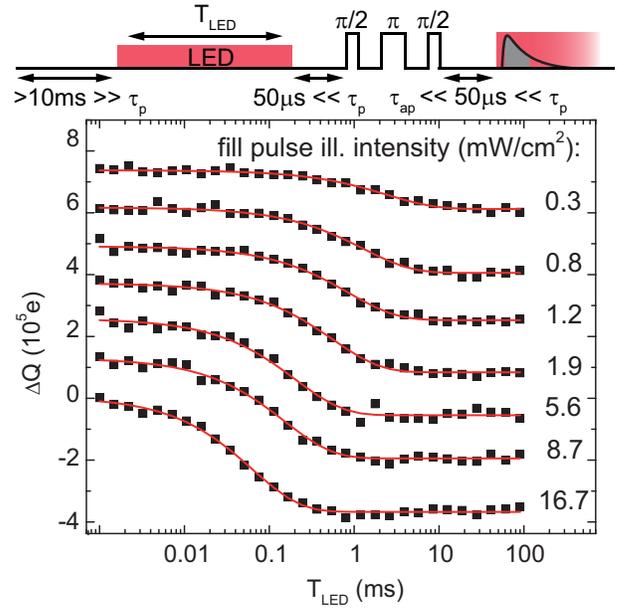} 
\par\end{centering}
\caption{\label{fig:FPL}
Pulse sequence to determine the generation time constant $\tau_\mathrm{g}$ with which new spin pairs are formed under illumination. Starting without illumination from $^{31}$P$^+$-P$_\mathrm{b0}^{-}$ spin pairs, a light pulse of length $T_\mathrm{LED}$ (fill pulse) creates new spin pairs $^{31}$P-P$_\mathrm{b0}$ by electron and hole capture from the conduction and valence band (cf. Fig.~\ref{fig:RateScheme}). The amount of newly created spin pairs is determined by measuring the amplitude $\Delta Q$ of a phase-cycled spin echo as a function of $T_\mathrm{LED}$. 
The results are shown in the lower part for different intensities of the fill light pulse (black squares) while the intensity of the detection light pulse is kept constant at 16.7~mW/cm$^2$ resulting in similar $\Delta Q$ for long $T_\mathrm{LED}$. $\Delta Q$ is shown as measured for the highest illumination intensity while all other traces are offset for clarity.
The data is fitted with a stretched exponential dependence with a characteristic time constant given by $2/(\frac{1}{\tau_\mathrm{g}}+\frac{2}{\tau_\mathrm{p}})$ (red lines).
}
\end{figure}

In the next set of experiments using pulsed illumination, we directly determine the generation rate of new spin pairs after a recombination process has taken place. To this end, we employ the pulse sequence sketched in Fig.~\ref{fig:FPL}. We switch off the illumination at least 10~ms before the pulse sequence, much longer than the recombination time constants of antiparallel and parallel spin pairs (cf. Sections~\ref{ss:rs} and~\ref{ss:rt}) to ensure that all spin pairs have recombined. At the beginning of the pulse sequence  
without illumination all spin pairs are therefore in the charged $^{31}$P$^+$-P$_\mathrm{b0}^{-}$ state. A light pulse of length $T_\mathrm{LED}$ (fill pulse) generates new spin pairs with a rate $1/\tau_\mathrm{g}$ depending on the intensity of the light pulse, which accumulate in the parallel state since $\tau_\mathrm{p}\gg\tau_\mathrm{ap}$. 
The amount of newly created spin pairs is determined by measuring the amplitude $\Delta Q$ of a phase-cycled spin echo as a function of $T_\mathrm{LED}$. 
Following the discussion in the Appendix~\ref{Appendix_FPL}, the echo amplitude is expected to increase exponentially with a characteristic time constant of 
\begin{equation}
\frac{1}{\lambda_1}=-\frac{2}{\left(\frac{1}{\tau_\mathrm{g}}+\frac{2}{\tau_\mathrm{p}}\right)}.
\end{equation}
This means that the time constant observed in the fill pulse length experiment is determined by the faster of the two processes, either the generation of new spin pairs or the recombination of parallel spin pairs. 

\begin{figure}[t!]
\noindent \begin{centering}
\includegraphics[width=8cm]{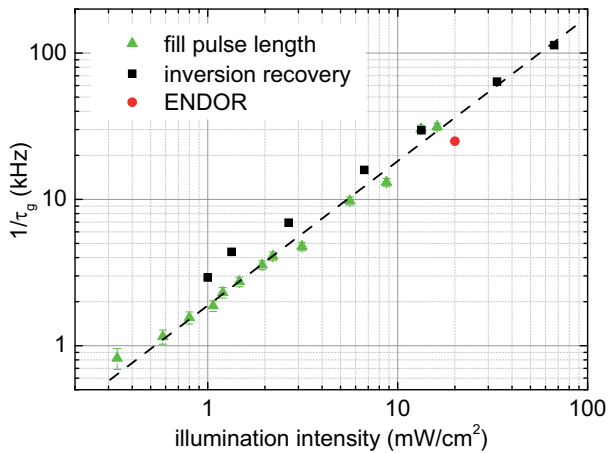} 
\par\end{centering}
\caption{\label{fig:g_summary}
Summary of the generation rates of new spin pairs $1/\tau_\mathrm{g}$ determined from (i) the variation of the fill pulse length (green triangles), (ii) the inversion recovery under continuous illumination (black squares) and (iii) the detection using the ionized donor nucleus (red dot) as a function of the illumination intensity. All experiments consistently show a linear dependence of the generation rate as a function of the photocurrent (dashed line). Assuming a constant mobility, this results in a linear dependence of the generation rate on the carrier density as expected for an electron capture process.
}
\end{figure}

The experimental results are shown in Fig.~\ref{fig:FPL} for different intensities of the first light pulse (black squares, offset for clarity) while the intensity of the detection light pulse is kept constant at 16.7~mW/cm$^2$.
Starting from $\Delta Q=0$ for short fill pulses, the absolute value of the echo amplitude increases for increasing $T$ with a characteristic time constant determined by fitting the data with a stretched exponential (red lines) of the form 
\begin{equation}
\Delta Q\propto e^{\mbox{\large$-\left(\frac{t}{2}\left(\frac{1}{\tau_\mathrm{g}}+\frac{2}{\tau_\mathrm{p}}\right)\right)^n$}},
\end{equation}
where the value of $\tau_\mathrm{p}$=1.2~ms is fixed as determined in Sections~\ref{ss:rs} and~\ref{ss:rt} and $n=0.8$. A summary of the obtained rates $1/\tau_\mathrm{g}$ is plotted in Fig.~\ref{fig:g_summary} (green triangles) together with the $1/\tau_\mathrm{g}$ values obtained by fitting the results of the inversion recovery experiment [Fig.~\ref{fig:InvRecIllu}(a)] under continuous illumination (black squares). Both experiments show consistent generation rates which increase linearly with increasing illumination intensity, confirming the assignment of $\tau_\text{ap}$ and $\tau_\text{g}$ in the previous Sect.~\ref{ss:rs}. Assuming a constant mobility, this results in a linear dependence of the generation rate on the carrier density as expected for an electron capture process~\cite{Loewenstein1966}. 
The fitting procedure of the fill pulse length dependence only gives meaningful results for $\tau_\mathrm{g}$ as long as $\tau_\mathrm{g}\lesssim \tau_\mathrm{p}/2$. Since $\tau_\mathrm{g}$ increases for decreasing illumination intensities, its value can not be determined at illumination intensities lower than 0.3~mW/cm$^2$ in our sample. 

\begin{figure}[t!]
\noindent \begin{centering}
\includegraphics[width=8cm]{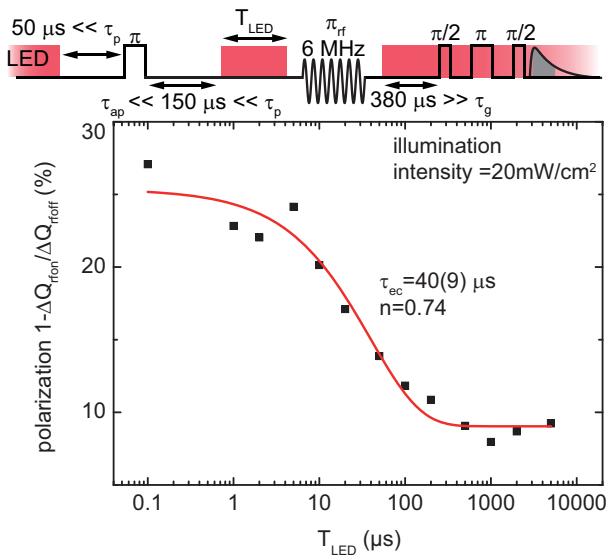} 
\par\end{centering}
\caption{\label{fig:ENDORg}
Pulse sequence to measure the electron capture rate using the nuclear spin of the ionized $^{31}$P donor as a probe. The first part of the pulse sequence removes the electron from the $^{31}$P donor for one orientation of its nuclear spin with respect to the $B_0$ field as described in Ref.~\cite{Dreher2012}, where the time interval of 150 $\mu$s is chosen such that the antiparallel spin pairs have recombined while the parallel spin pairs have not. During the following light pulse of length $T_\mathrm{LED}$ the $^{31}$P$^+$ captures electrons with a rate $1/\tau_\mathrm{ec}$. The number of ionized $^{31}$P donors after the light pulse is probed by applying an rf $\pi$ pulse with a frequency corresponding to the Larmor frequency of the $^{31}$P$^+$ nuclear spin. The resulting nuclear polarization determined by a detection spin echo reflects the number of $^{31}$P$^+$ nuclear spins before the rf pulse.
In the lower part of the figure, the polarization is plotted as a function of the length $T_\mathrm{LED}$ of the light pulse with an illumination intensity of 20~mW/cm$^2$ (black squares). The polarization decays with a time constant of $\tau_\mathrm{ec}$=40~$\mu$s as determined by a stretched exponential fit (red line).}
\end{figure}

The experiments described so far allow us to determine the generation rate of new spin pairs. Referring to Fig.~\ref{fig:RateScheme}, two processes, namely the capture of an electron by the $^{31}$P$^+$ and the capture of a hole by the P$_\mathrm{b0}^-$, are involved in the generation of new spin pairs. To decide whether the observed time constant is determined by the electron or hole capture process, we can use the nuclear spin of the ionized $^{31}$P donor following the approach described in Ref.~\cite{Dreher2012}. The corresponding pulse sequence is sketched in Fig.~\ref{fig:ENDORg}.

Starting from the steady state under illumination, in a first step we selectively depopulate the $^{31}$P donors associated with one orientation of their nuclear spin with respect to the $B_0$ field, e.g., spin up. This is done by switching off the LED and applying a $\pi$ pulse on the corresponding $^{31}$P electron spin transition. The spin pairs with $^{31}$P nuclear spin up are now in an antiparallel state and therefore recombine with a time constant of $\tau_\mathrm{ap}\approx$15~$\mu$s while the spin pairs with the $^{31}$P nuclear spin down remain stable on the much longer timescale $\tau_\mathrm{p}\approx$1.2~ms. This results in ionized $^{31}$P donors with one preferred direction of their nuclear spin since without illumination no new spin pairs are generated. The nuclear spins of the ionized $^{31}$P can be resonantly flipped by applying a rf $\pi$ pulse with a frequency of 6.036~MHz corresponding to the Larmor frequency of the $^{31}$P$^+$ nuclear spin at B$_0$=350.6~mT~\cite{Dreher2012}. This results in a polarization of the nuclear spins which can be detected by comparing the amplitudes of spin echoes for resonant $\Delta Q_\mathrm{rfon}$ and off-resonant $\Delta Q_\mathrm{rfoff}$ rf pulses measured after repopulating the donors with a light pulse. The polarization is then given by 1-$\Delta Q_\mathrm{rfon}$/$\Delta Q_\mathrm{rfoff}$. Introduction of a light pulse of length $T_\mathrm{LED}$ between the depopulation pulse and the rf pulse repopulates the $^{31}$P$^+$ ~donors with a rate $1/\tau_\mathrm{ec}$, thereby reducing the achievable nuclear spin polarization. Figure~\ref{fig:ENDORg} shows the polarization for an illumination intensity of 20~mW/cm$^2$ as a function of the light pulse length $T_\mathrm{LED}$ (black squares). The polarization decays with a time constant of $\tau_\mathrm{ec}$=40(9)~$\mu$s and an exponent of $n$=0.5 as determined by a stretched exponential fit (red line).

The rate for the generation of new $^{31}$P-P$_\mathrm{b0}$ spin pairs, however, is given by $1/\tau_\mathrm{g}=1/\tau_\mathrm{ec}+1/\tau_\mathrm{hc}$ assuming that the electron and hole capture processes are uncorrelated. Since the time constant observed in the ENDOR experiment, which directly measures the electron capture time constant $\tau_\mathrm{ec}$, fits well onto the dependence observed for $1/\tau_\mathrm{g}$ by both experiments so far (Fig.~\ref{fig:g_summary}), we associate $\tau_\mathrm{g}=\tau_\mathrm{ec}$. 
To access $\tau_\mathrm{hc}$, a similar ENDOR experiment could be performed using the hyperfine coupling of the P$_\mathrm{b0}$ to a nearby $^{29}$Si nuclear spin~\cite{Stesmans98}. For the P$_\mathrm{b0}$ in the negative charge state, the nearby $^{29}$Si nuclear spins should be polarizable by an rf pulse at their free Larmor frequency, while for a neutral P$_\mathrm{b0}$ the nuclear spin transition frequency is changed by the hyperfine interaction.

The fact that the polarization does not decay to zero indicates a steady state nuclear polarization created by the illumination as has been observed in illuminated phosphorus-doped silicon at higher (8~T) magnetic fields~\cite{Mccamey2009}. The reason for this steady state polarization appears to be related to the excitation of carriers into the conduction band. This, however, requires further experiments for clarification and is beyond the scope of this investigation.

The capture rate $1/\tau_\mathrm{ec}$ can be calculated from the capture rate constant $\sigma_\mathrm{ec}$ of $^{31}$P$^+$, which has been determined already in several studies, notably also using electron spin resonance techniques~\cite{Levitt1961,Loewenstein1966,Norton1973}, allowing the comparison of the experimental results presented here with $1/\tau_\mathrm{ec}$=$\sigma_\mathrm{ec}\cdot n_\mathrm{e}$, where $n_\mathrm{e}$ denotes the carrier density in the conduction band.
We estimate the value of $n_\mathrm{e}$ in our sample for the illumination intensity of $20$~mW/cm$^2$ used in the ENDOR experiment. For red light with a wavelength of 635~nm, this corresponds to a photon flux of $I_\mathrm{photon}$=6.4$\cdot 10^{16}$~cm$^{-2}$s$^{-1}$ incident on the sample surface. With an absorption coefficient of $\alpha_\mathrm{Si}\approx$2$\cdot 10^3$~cm$^{-1}$ (77~K)~\cite{Dash1955} and a reflectivity of $\sim$40\%~\cite{Koynov2006}, a fraction of $0.4\cdot e^{-\alpha_\text{Si}\cdot d}\approx$0.4 of the incident photons are absorbed in the $d$=2.5~$\mu$m thick device layer of the SOI sample. We expect that the optically excited carriers rapidly diffuse within this layer, so that we can assume a spatially homogeneous carrier generation rate of $G=0.4\cdot I_\mathrm{photon}/d=1.0\cdot 10^{20}$~cm$^{-3}$s$^{-1}$. The carrier density is then given by $n_\mathrm{e}=G\cdot \tau_\mathrm{l}$, where $\tau_\mathrm{l}$ denotes the carrier lifetime. We experimentally determine an upper bound for $\tau_\mathrm{l}$ by measuring the decay time constant of the current transient after a rectangular illumination pulse and find a value of 100~ns, which, however, corresponds to the bandwidth limit of our measurement setup. In silicon, surface recombination velocities of $S=10^4$~cm/s have been reported for the Si/SiO$_2$ interface with a native oxide~\cite{Lehner2003} corresponding to a carrier lifetime of $\tau_\mathrm{l}=d/2S$=10~ns, where the factor of 2 accounts for recombination at the surface and the buried oxide. Using this value of $\tau_\mathrm{l}$ rather than the above upper bound results in $n_\mathrm{e}$=10$^{12}$~cm$^{-3}$.

References~\cite{Loewenstein1966}and~\cite{Levitt1961} report values of $\sigma_\mathrm{ec}$=7x10$^{-6}$~cm$^{3}$s$^{-1}$ at 4.2~K and $\sigma_\mathrm{ec}$=5x10$^{-6}$~cm$^{3}$s$^{-1}$ at 5~K, respectively, while an one order of magnitude higher value of $\sigma_\mathrm{ec}$=6x10$^{-5}$~cm$^{3}$s$^{-1}$ at 5~K is found in Ref.~\cite{Norton1973}. Taking the lower value of $\sigma_\mathrm{ec}$=5x10$^{-6}$~cm$^{3}$s$^{-1}$, we expect an electron capture rate of $1/\tau_\mathrm{ec}$=$\sigma_\mathrm{ec}\cdot n_\mathrm{e}$=5x10$^6$s$^{-1}$, which is a factor of 10 larger than the values experimentally observed here at 20~mW/cm$^2$.
Most importantly, the method presented here allows to determine the electron capture rate directly for $^{31}$P$^+$ donors at the Si/SiO$_2$ interface. The remaining difference to the published data for $^{31}$P$^+$ in bulk Si might be attributable to higher surface recombination velocities $S$ at the Si/Au contacts, since Au acts as a very efficient recombination center~\cite{Sze2001} or too large local electric fields near the contacts which would reduce the capture rate constant~\cite{Sclar1984}.

An alternative approach of measuring the generation rate of new spin pairs in amorphous hydrogenated silicon has been presented in Ref.~\cite{Behrends09}. In this approach, the separation of spin pair recombination and generation rates is achieved by using a rotary echo.

\subsection{Recombination Rate of Parallel Spin Pairs}
\label{ss:rt}

\begin{figure}[t!]
\begin{centering}
\includegraphics[width=8cm]{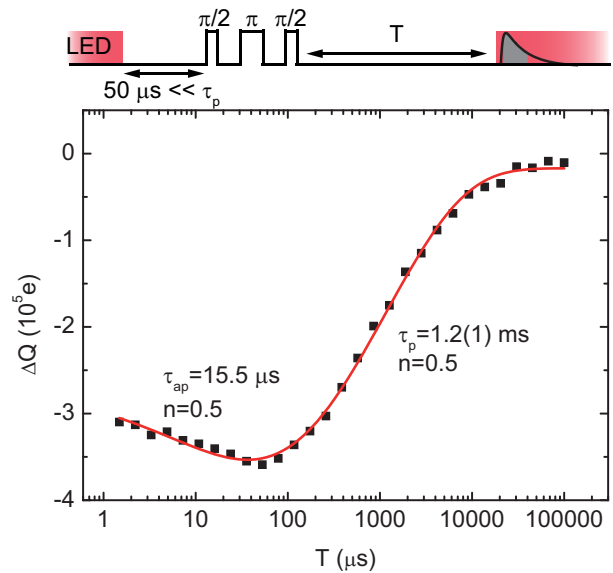} 
\par\end{centering}
\caption{\label{fig:DPS}
Pulse sequence to determine the recombination time of parallel spin pairs $\tau_\mathrm{p}$. 
The integrated current transient is recorded as a function of the time interval $T$ between the detection spin echo and the detection light pulse. The experimentally observed decay (black symbols) can be fitted with two stretched exponentials (red line) with an exponent of 0.5 and time constants of 15.5~$\mu$s and 1.2(1)~ms, respectively .
}
\end{figure}

The recombination time $\tau_\mathrm{p}$ of parallel spin pairs, which has been observed in the inversion recovery measurement shown in Fig.~\ref{fig:InvRecDark}, can be determined directly by measuring the amplitude of a spin echo with pulsed optical excitation as a function of the time $T$ between the spin echo and the detection light pulse (see Fig.~\ref{fig:DPS}). Starting from a steady state with only parallel spin pairs, a spin echo creates antiparallel spin pairs for the (-x) phase and parallel spin pairs for the (+x) phase of the last $\pi$/2 pulse. For waiting times $\tau_\mathrm{ap}\ll T \ll \tau_\mathrm{p}$ the antiparallel spin pairs have recombined while the parallel spin pairs have not. Switching on the illumination results in a current transient with an amplitude proportional to the number of recombined spin pairs. For $T>\tau_\mathrm{p}$ the parallel spin pairs recombine as well, decreasing the contrast between the (+x) and (-x) phases of the spin echo and thus the observed spin echo amplitude. 

Figure~\ref{fig:DPS} shows the amplitude of the detection echo $\Delta Q\propto n_\mathrm{ap}-n_\mathrm{p}$ (black symbols) as a function of the waiting time $T$ between the spin echo and the detection light pulse. The observed decay can again be fitted by two stretched exponentials with an exponent of 0.5 and time constants of 15.5~$\mu$s and 1.2(1)~ms, respectively. The former reflects the recombination rate of antiparallel spin pairs as determined by the inversion recovery experiment (Fig.~\ref{fig:InvRecDark}), while the latter is attributed to the decay of parallel spin pairs. For the fit (red line) the exponent and time constant of the first decay have been fixed to the values determined by the inversion recovery experiment.

\begin{figure}[t!]
\begin{centering}
\includegraphics[width=8cm]{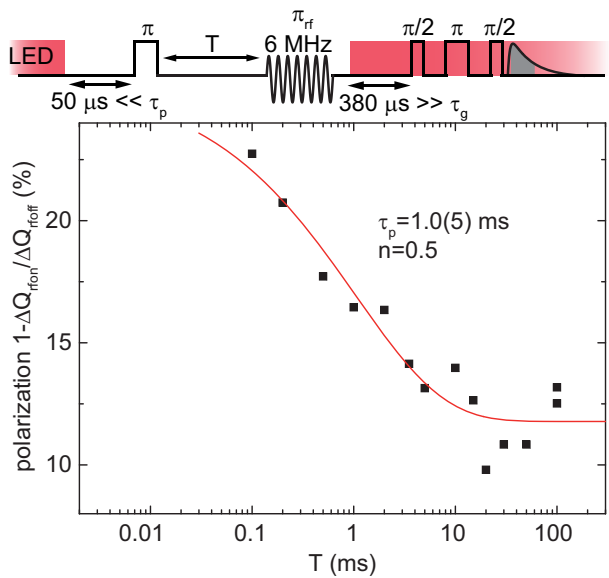} 
\par\end{centering}
\caption{\label{fig:ENDORrt}
Pulse sequence to measure the recombination rate of antiparallel spin pairs using the nuclear spin of the ionized $^{31}$P donor as a probe. First, the $^{31}$P donors with one orientation of their nuclear spins with respect to the magnetic field are depopulated. Application of an rf $\pi$ pulse on the $^{31}$P$^+$ nuclear spin transition after a waiting time $T$ creates a nuclear spin polarization which is measured using a spin echo after repopulating the donors with a light pulse. The nuclear spin polarization (black squares) decays for large time intervals $T$ due to the recombination of parallel spin pairs. The decay can be fitted with a stretched exponential (red line) with $\tau_\mathrm{p}$=1.0(5)~ms and $n$=0.5.
}
\end{figure}

We can also use the nuclear spin of the ionized $^{31}$P donor to measure the recombination time of parallel spin pairs similar to the measurement of the generation rate shown in Fig.~\ref{fig:ENDORg}. Again, we first selectively depopulate one hyperfine transition of the $^{31}$P donors, e.g., with nuclear spin up. We then apply a $\pi$ pulse on the nuclear spin transition of the ionized donor to create a nuclear spin polarization which is measured using a spin echo similar to the experiment described in Sect.~\ref{ss:generation}. For large waiting times $T$ between the depopulation sequence and the rf pulse also the parallel spin pairs start to recombine so that also the $^{31}$P with nuclear spin down become ionized, thereby reducing the polarization created by the rf pulse. Figure~\ref{fig:ENDORrt}(a) shows the polarization $1-\Delta Q_\mathrm{rfon}/\Delta Q_\mathrm{rfoff}$ as a function of $T$ which decays with a stretched exponential with a time constant of $\tau_\mathrm{p}$=1.0(5)~ms and an exponent $n$=0.5. This is in good agreement with $\tau_\mathrm{p}$=1.2~ms obtained in the experiment described in Fig.~\ref{fig:DPS}. Again, for large $T$ the polarization does not decay to zero as observed in Fig.~\ref{fig:ENDORg}.

The recombination of parallel spin pairs can be interpreted as a two step process consisting of a spin flip of either the $^{31}$P or the P$_\mathrm{b0}$ spin and a recombination of antiparallel spin pairs. A value of $\approx$10~s has been determined for the spin relaxation rate for $^{31}$P electron spins in bulk silicon at 5~K~\cite{FeherII59Relaxation} and comparable $^{31}$P concentrations, much larger than the value of $\tau_\mathrm{p}$=1.2~ms observed in our experiments. The spin relaxation time of dangling bonds at the Si/SiO$_2$-interface in crystalline silicon has so far not been studied experimentally. There are, however, detailed studies of the dangling bond relaxation in amorphous silicon which report relaxation times of 0.1-1~ms at 0.3~T and 5~K~\cite{Stutzmann1983, Askew86}. Since these values are comparable with $\tau_\mathrm{p}$=1.2~ms obtained in this work, we tentatively attribute the recombination of parallel spin pairs to be caused by the spin relaxation of the P$_\mathrm{b0}$ spins.

\section{Pulsed EDMR Photocurrent Transients}
\label{ss:transients}
The time constants involved in the recombination process as analyzed so far also determine the time-dependence of the spin-dependent part of the photocurrent transient after resonant excitation by a short microwave pulse~\cite{Boehme03EDMR}.
In this section, we set up a rate equation model describing this pulsed EDMR current transient and compare the simulated current transients with experimental results. For this, we extend the model described in Sect.~\ref{ss:rateequations} by including the hole capture time constant $\tau_\mathrm{hc}$ explicitly.

\begin{figure}[t!]
\begin{centering}
\includegraphics[width=8cm]{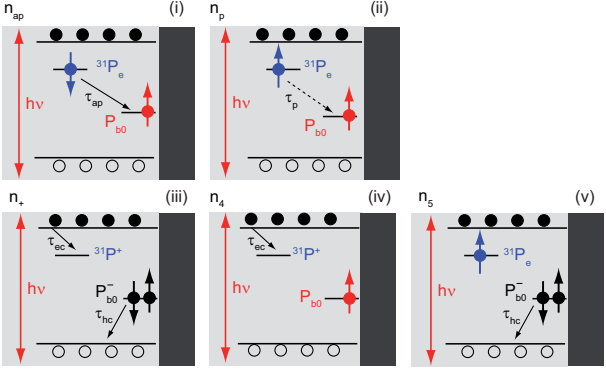}
\par\end{centering}
\caption{\label{Fig_time constants_RateScheme}
Dynamics of the populations of the spin states. These include the parallel (i) and antiparallel states (ii) of the spin pair, the $^{31}$P$^+$-P$_\mathrm{b0}^-$ state (iii), the $^{31}$P$^+$-P$_\mathrm{b0}$ state (iv), and the $^{31}$P-P$_\mathrm{b0}^-$ state (v). The corresponding populations are denoted by $n_\text{p}$, $n_\text{ap}$, $n_\text{+}$, $n_\text{4}$, and $n_\text{5}$, respectively.
}
\end{figure}

We discuss the dynamics of the populations of the spin states in terms of the five states sketched in Fig.~\ref{Fig_time constants_RateScheme}. This includes the parallel (i) and antiparallel states (ii) of the spin pair, the $^{31}$P$^+$-P$_\mathrm{b0}^-$ state (iii), the $^{31}$P$^+$-P$_\mathrm{b0}$ state (iv), and the $^{31}$P-P$_\mathrm{b0}^-$ state (v). The corresponding populations are denoted by $n_\text{p}$, $n_\text{ap}$, $n_\text{+}$, $n_\text{4}$, and $n_\text{5}$, respectively. These five states represent all combinations of $^{31}$P, $^{31}$P$^+$ as well as P$_\mathrm{b0}$ and P$_\mathrm{b0}^-$ and, therefore, they are normalized such that $n_\text{p}+n_\text{ap}+n_\text{+}+n_\text{4}+n_\text{5}=1$, so that, e.g., $n_\text{ap}$ denotes the fraction of $^{31}$P-P$_\mathrm{b0}$ pairs in the antiparallel configuration. 

Analogous to Sect.~\ref{ss:rateequations}, the time evolution of the state vector $\rho_5$ is governed  by the differential equation
\begin{equation}
\label{eq:time constants_rateequations}
\frac{\text{d}}{\text{d}t}\rho_5=\tilde{R_5}\rho_5,
\end{equation}
with
\begin{equation}
\rho_5=
\begin{pmatrix}
n_\mathrm{p} \\
n_\mathrm{ap} \\
n_\mathrm{+} \\
n_\mathrm{4} \\
n_\mathrm{5} \\
\end{pmatrix},
\end{equation}
and
\begin{equation}
\label{eq:time constants_recombmatrix}
\tilde{R_5}=
\begin{pmatrix}
-1/\tau_\mathrm{p} & 0 & 0 & 1/2\tau_\mathrm{ec} & 1/2\tau_\mathrm{hc}\\
0 & -1/\tau_\mathrm{ap} & 0 & 1/2\tau_\mathrm{ec} & 1/2\tau_\mathrm{hc}\\
1/\tau_\mathrm{p} & 1/\tau_\mathrm{ap} & -1/\tau_\mathrm{ec}-1/\tau_\mathrm{hc} & 0 & 0\\
0 & 0 & 1/\tau_\mathrm{hc} & -1/\tau_\mathrm{ec} & 0 \\
0 & 0 & 1/\tau_\mathrm{ec} & 0 & -1/\tau_\mathrm{hc}\\
\end{pmatrix}.
\end{equation}
The solution of Eq.~\eqref{eq:time constants_recombmatrix} is given by
\begin{equation}
\label{eq:time constants_rhoevolution}
\rho_5(t)=\rho_5(0)\cdot\text{e}^{\tilde{R_5}\cdot t}
\end{equation}
describing the time evolution of $\rho_5(t)$.
The steady state vector $\rho_\text{5eq}$ is determined by the condition
\begin{equation}
\tilde{R_5}\cdot \rho_\text{5eq}=\tilde{R_5}\begin{pmatrix}
n_\mathrm{p}^\mathrm{eq} \\
n_\mathrm{ap}^\mathrm{eq} \\
n_\mathrm{+}^\mathrm{eq} \\
n_\mathrm{4}^\mathrm{eq} \\
n_\mathrm{5}^\mathrm{eq} \\
\end{pmatrix}=0.
\end{equation}

Having established the dynamics of the spin state populations, we calculate the changes of the carrier densities in the conduction and valence bands $n_\text{e}$ and $n_\text{h}$, respectively.
Assuming that at low temperatures only photo-excited carriers are present in the conduction and valence bands, the time dependence of the carrier densities are given by
\begin{equation}
\label{eq:appendix_carrierdensity}
\begin{split}
\frac{\text{d}n_\text{e}}{\text{d}t}&=G-\frac{n_\text{e}(t)}{\tau_\text{l}}-\frac{n_\text{sp}}{\tau_\text{ec}}(n_+(t)+n_4(t))\\
\frac{\text{d}n_\text{h}}{\text{d}t}&=G-\frac{n_\text{h}(t)}{\tau_\text{l}}-\frac{n_\text{sp}}{\tau_\text{hc}}(n_+(t)+n_5(t)),
\end{split}
\end{equation}
where $G$ is the generation rate of electron-hole pairs, $\tau_\text{l}$ the carrier lifetime in the sample assuming monomolecular spin-independent recombination as confirmed in our sample by the linear dependence of the photocurrent on the illumination intensity, and $n_\text{sp}$ the density of $^{31}$P-P$_\mathrm{b0}$ pairs, so that, e.g., $n_\text{sp}\cdot n_\text{ap}$ denotes the total density of antiparallel spin pairs. The third terms in the Eqs.~\eqref{eq:appendix_carrierdensity} describe the change of the carrier densities caused by spin pair recombination. The electron and hole capture rates $1/\tau_\mathrm{ec}$ and $1/\tau_\mathrm{hc}$ are proportional to $n_\text{e}(t)$ and $n_\text{h}(t)$, but we neglect the resulting implicit time dependence of $\tau_\mathrm{ec}$ and $\tau_\mathrm{hc}$, since the third terms in Eqs.~\eqref{eq:appendix_carrierdensity} are small compared with the first two terms. This appears justified since the relative current changes detected by EDMR are usually \textless 10$^{-2}$ so that the resulting variations in $n_\text{e}(t)$ and $n_\text{h}(t)$ are negligible compared with their steady state values. 

With these assumptions, we calculate the steady state of Eqs.~\eqref{eq:appendix_carrierdensity} $\mathrm{d}n_\mathrm{e/h}/\mathrm{dt}=0$ given by
\begin{equation}
\begin{split}
n_\text{e}(t)&=\tau_\text{l}\cdot G-\frac{n_\text{sp}\cdot \tau_\text{l}}{\tau_\text{ec}}\cdot(n_+(t)+n_4(t))\\
n_\text{h}(t)&=\tau_\text{l}\cdot G-\frac{n_\text{sp}\cdot \tau_\text{l}}{\tau_\text{hc}}\cdot(n_+(t)+n_5(t)).
\end{split}
\end{equation}
We hereby also take into account that the carrier lifetime $\tau_\text{l}$ is short compared to the characteristic time constants of the spin pair, so that $n_\text{e}(t)$ and $n_\text{h}(t)$ instantaneously follow the time-dependence of $\rho_5(t)$. In Sect.~\ref{ss:generation}, we determined an upper bound of $\tau_\text{l}$=100~ns much shorter than the shortest time constant of the spin pair $\tau_\text{ap}\approx$2~$\mu$s.

The change of the carrier densities after a resonant microwave pulse is then given by
\begin{equation}
\begin{split}
\Delta n_\text{e}(t)&=\frac{n_\text{sp}\cdot \tau_\text{l}}{\tau_\text{ec}}\cdot(\Delta n_+(t)+\Delta n_4(t))\\
\Delta n_\text{h}(t)&=\frac{n_\text{sp}\cdot \tau_\text{l}}{\tau_\text{hc}}\cdot(\Delta n_+(t)+\Delta n_5(t)),
\end{split}
\end{equation}
where $\Delta n_{i}=n_{i}^\text{eq}-n_{i}(t)$ with $i=+,4,5$.
This results in a photoconductivity change of 
\begin{equation}
\Delta \sigma=e\left[\mu_\text{e}\Delta n_\text{e}+\mu_\text{h}\Delta n_\text{h}\right],
\end{equation}
where $\mu_\text{e}$ and $\mu_\text{h}$ denote the electron and hole mobilities, respectively.
With this, the relative change in photoconductivity becomes
\begin{equation}
\label{eq:time constants_Deltasigma}
\begin{split}
\frac{\Delta \sigma}{\sigma} & =\frac{e\left(\mu_\text{e}\Delta n_\text{e}+\mu_\text{h}\Delta n_\text{h}\right)}{e\cdot G\cdot \tau_\text{l}\cdot(\mu_\text{e}+\mu_\text{h})}\\
& =\frac{n_\text{sp}}{G(1+\gamma)}\cdot\\
& \cdot \left[\frac{\gamma}{\tau_\text{ec}}(\Delta n_+(t)+\Delta n_4(t))+\frac{1}{\tau_\text{hc}}(\Delta n_+(t)+\Delta n_5(t))\right],
\end{split}
\end{equation}
where we have introduced the ratio of the electron and hole mobilities $\gamma=\mu_\text{e}/\mu_\text{h}$.

We will briefly discuss some implications of Eq.~\eqref{eq:time constants_Deltasigma}. First, it predicts that the maximum relative change of the photocurrent $\Delta I/I \propto \Delta \sigma/\sigma$ only weakly depends on the illumination intensity. The electron and hole capture rates $1/\tau_\text{ec}$ and $1/\tau_\text{hc}$ are proportional to the carrier density in the conduction and valence band, respectively. Therefore, both capture rates as well as the electron-hole pair generation rate $G$ are proportional to the illumination intensity, so that their ratio in Eq.~\eqref{eq:time constants_Deltasigma} is independent of the illumination intensity. Only the $\Delta n_i(t)$ depend on the illumination intensity via $\tau_\text{ec}$ and $\tau_\text{hc}$, with, however, only small variations as long as $\tau_\text{ec},\tau_\text{hc}>\tau_\text{ap}$ as can be confirmed by numerical simulations of Eq.~\eqref{eq:time constants_rhoevolution}. 
\begin{figure}[t!]
\begin{centering}
\includegraphics[width=8cm]{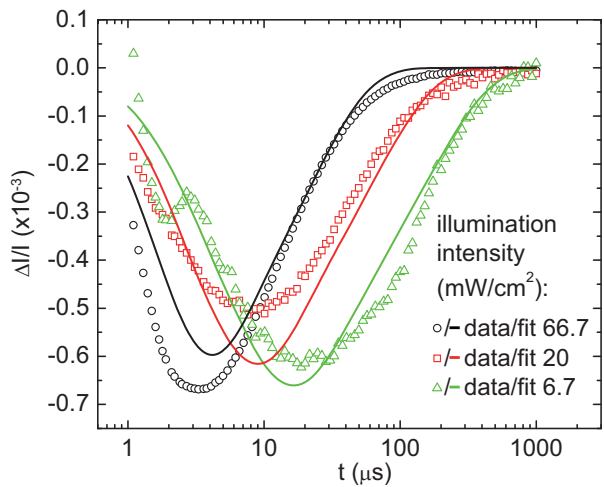} 
\par\end{centering}
\caption{\label{fig:transients}
Spin-dependent photocurrent transients normalized to the photocurrent after resonant excitation by a short microwave $\pi$ pulse for different illumination intensities (open symbols). The solid lines show fits of the experimental data based on Eq.~\eqref{eq:time constants_Deltasigma} with fitting parameters as summarized in Table~\ref{fig:time constants_Tablefit}.
}
\end{figure}

To compare the predictions of Eq.~\eqref{eq:time constants_Deltasigma} with experimental data, we
show photocurrent transients recorded for three different illumination intensities in Fig.~\ref{fig:transients}. We use a Femto current amplifier with a bandwidth of $\approx$1~MHz without any additional filtering for this experiment, since in particular high-pass filtering alters the shape of the current transient. The pulse used for resonant excitation has a length of 120~ns with a microwave magnetic field amplitude of 0.15~mT corresponding to a $\pi$ pulse. The non-resonant part of the photocurrent transient was removed by recording the photocurrent transient at two additional values of the static magnetic field where no resonances are observed and subtracting the linearly interpolated transient as a background~\cite{Stegner06}. 
The resonant part of the current transients normalized to the photocurrent is shown in Fig.~\ref{fig:transients} (open symbols). All three transients show a similar maximum value of $\Delta I/I$, although the illumination intensity varies over one order of magnitude, confirming the prediction of Eq.~\eqref{eq:time constants_Deltasigma}. Similar observations have been reported in Ref.~\cite{Lehner2003} for cw~EDMR measurements in differently passivated silicon. The rise time increases with decreasing illumination intensity, demonstrating that it is not limited by the bandwidth of the current measurement setup. It is rather determined by the recombination time of antiparallel spin pairs in agreement with the illumination dependence observed in Fig.~\ref{fig:InvRecIllu}(b). We also observe that the decay time constants of the current transient increase with decreasing illumination intensity, indicating that they are related to the electron and hole capture time constants. This assignment is in agreement with the physical picture that a recombination process has to occur before a change in the photocurrent is observable, while the electron and hole capture processes restore the steady state causing the photocurrent transient to decay.    
\begin{table}[ttt]
\centering
\caption{
Summary of the parameters used for the fits of the photocurrent transients in Fig.~\ref{fig:transients}. Only two independent fitting parameters are used: the $^{31}$P-P$_\mathrm{b0}$ density $n_\text{sp}=4\cdot 10^{12}$~$\frac{1}{\text{cm}^3}$ and the ratio $\tau_\text{ec}/\tau_\text{hc}\approx 2$. The value of $\gamma=\mu_\text{e}/\mu_\text{h}$=2 is estimated based on Ref.~\cite{Morin1954} and all other parameters are extracted from the results obtained in Sect.~\ref{results}.} 
\label{fig:time constants_Tablefit}%
\vspace*{0.2cm}
\begin{tabular}{c|c|c|c|c|c}
\hline \hline 
illumination & $G$ ($\frac{1}{\text{cm}^3\text{s}}$) & $\tau_\text{ap}$ ($\mu$s) & $\tau_\text{p}$ ($\mu$s) & $\tau_\text{ec}$ ($\mu$s) & $\tau_\text{hc}$ ($\mu$s)  \\
intensity ($\frac{\text{mW}}{\text{cm}^2}$) & & & & & \\
\hline
 6.7 & 5$\cdot 10^{19}$&7 & 1200 &90 & 45  \\
 20 & 1.6$\cdot 10^{20}$ & 4 & 1200 &30 & 15   \\
 66.7 & 5$\cdot 10^{20}$ & 2 & 1200 &9 & 5   \\
\hline \hline
\end{tabular}
\end{table}

For a more detailed comparison, we fit the three experimental photocurrent transients with the recombination model described by Eq.~\eqref{eq:time constants_recombmatrix} and Eq.~\eqref{eq:time constants_rhoevolution} in combination with Eq.~\eqref{eq:time constants_Deltasigma}. To this end, starting from $\rho_\text{5eq}$, we calculate $\rho_5(t)$ after an ideal $\pi$ pulse which exchanges the populations of the antiparallel and parallel spin states. We use only two free parameters to fit the data for all three illumination intensities: the density of $^{31}$P-P$_\mathrm{b0}$ pairs $n_\text{sp}$ and the ratio of the electron and hole capture time constants $\tau_\text{ec}/\tau_\text{hc}$. The recombination time constants of antiparallel and parallel spin pairs $\tau_\text{ap}$ and $\tau_\text{p}$ are extracted from Fig.~\ref{fig:InvRecIllu}(b) and Fig.~\ref{fig:DPS}, the electron capture rate is taken from Fig.~\ref{fig:g_summary}, the electron-hole pair generation rate $G$ is calculated from the measured illumination intensity, and the ratio of the electron and hole mobilities $\gamma$ is estimated to be $\sim$2~\cite{Morin1954}. All parameters thus determined are summarized in Table~\ref{fig:time constants_Tablefit}.
The best fits (solid lines in Fig.~\ref{fig:transients}) are obtained for $\tau_\text{ec}/\tau_\text{hc}$=5 and a $^{31}$P-P$_\mathrm{b0}$ pair density of $n_\text{sp}$=4$\cdot 10^{12}$~cm$^{-3}$ corresponding to an absolute number of 4$\cdot 10^7$ spin pairs in the sample. The fits reproduce the general features of the experimental data such as the near constant amplitude of the current transient for different illumination intensities and the increase of the decay time constant with decreasing illumination intensity. Some differences are however found, e.g., in the time constant of the decay in particular for the lowest illumination intensity. This shows that the rate equation model neglects some features, most of all the distribution of recombination time constants leading to the experimentally observed stretched exponential decays but also, e.g., the additional feature observed at $t$=2~$\mu$s for the lowest illumination intensity in Fig.~\ref{fig:transients}.

The value of $n_\text{sp}$=4$\cdot 10^{12}$~cm$^{-3}$ obtained from the fit can be compared with an estimation of the spin pair density $n_\text{est}$ based on the $^{31}$P concentration of 3$\cdot 10^{16}$~cm$^{-3}$ in the doped epilayer and the geometry of the sample. Since the resulting area density of $^{31}$P ~donors (6$\cdot 10^{10}$~cm$^{-2}$) is much smaller than the P$_\mathrm{b0}$ density ($\sim 1\cdot 10^{12}$~cm$^{-2}$~\cite{Stesmans98dbDensity}), the spin pair density is limited by the donors rather than by the P$_\mathrm{b0}$ centers, so that $n_\text{est}$=3$\cdot 10^{16}$~cm$^{-3}$.
This value is further reduced since we estimate that only $^{31}$P ~donors in a $\sim$6~nm part of the sample are observed~\cite{Suckert2013}, while the whole silicon layer above the buried oxide is illuminated and therefore contributes to the observed current. This reduces the density of $^{31}$P-P$_\mathrm{b0}$ pairs with respect to the whole volume through which the current flows to $n_\text{est}$=7$\cdot 10^{13}$~cm$^{-3}$, only a factor of 20 larger than the fit result of $n_\text{sp}$=4$\cdot 10^{12}$~cm$^{-3}$. However, in this estimate we again neglect the distribution of recombination rates within the spin pair ensemble, possibly explaining the observed discrepancy. 

The recombination model discussed here differs in several aspects from the model used, e.g., in Ref.~\cite{Boehme03EDMR} to describe the photocurrent transient in pulsed EDMR. First, we simplify the rate equations by neglecting the $^{31}$P-P$_\mathrm{b0}$ coupling of \textless 1~MHz~\cite{Suckert2013}, since it is much smaller than the $^{31}$P-P$_\mathrm{b0}$ Larmor frequency difference ($\approx$ 30~MHz), as well as the possibility of spin pair dissociation by emission of an electron (hole) into the conduction (valence) band characteristic for the KSM-model. Most importantly, we explicitly take the different electron and hole capture time constants into account, which have been neglected in the description of the time dependence of the spin pair dynamics in Ref.~\cite{Boehme03EDMR}. For the experimental conditions in this work, these time constants appear to be longer than the recombination time constant of antiparallel spin pairs, so that the time dependence of the decay of the photocurrent transient is mainly determined by $\tau_\text{ec}$ and $\tau_\text{hc}$, while the rise time is determined by $\tau_\text{ap}$ as confirmed by the transients shown in Fig.~\ref{fig:transients}. For much higher illumination intensities or in different samples, $\tau_\text{ec}$ and $\tau_\text{hc}$ eventually can become shorter than $\tau_\text{ap}$, in which case the decay time constant would be given by $\tau_\text{ap}$ as described in Ref.~\cite{Boehme03EDMR}, while the rise time would be determined by $\tau_\text{ec}$ and $\tau_\text{hc}$ or the bandwidth of the detection setup.


\section{Implications of a Broad Distribution of Recombination Time constants}
\label{ss:distribution}

In most pulsed EDMR experiments, boxcar integration of the spin-dependent part of the photocurrent transient is used to determine the state of the spin pair after a pulse sequence~\cite{Boehme03EDMR,Stegner06}. As we have seen in the preceding section, the time dependence of the photocurrent transient itself depends on the spin pair recombination and formation time constants, so that the results of measurements of these time constants are expected to depend on the chosen boxcar integration interval. In this section, we will show exemplarily using a simplified model for the current transient and a distribution of antiparallel recombination rates, that the characteristic time constant measured by an inversion recovery experiment is determined solely by the boxcar integration interval. We further compare the results predicted by this model with measurements of $\tau_\mathrm{ap}$ for different boxcar integration intervals.  


To allow for a simple description of the current transient, we consider the case of high illumination intensities, so that the electron and hole capture time constants are much smaller than the recombination time constants of antiparallel spin pairs. In this case, which corresponds to the situation discussed in Ref.~\cite{Boehme03EDMR}, the characteristic decay time constant of the photocurrent transient is equal to $\tau_\mathrm{ap}$. For longer capture time constants, a reasoning similar to the argument developed now can be applied to measurements of the electron and hole capture time constants.
Under these conditions, the current transient is given by
\begin{equation}
\label{eq:time constants_transient}
I(t)\propto r\cdot \text{e}^{-r\cdot t},
\end{equation}
where we have introduced the recombination rate of antiparallel spin pairs $r=1/\tau_\text{ap}$. The additional factor of $r$ before the exponential takes into account that the recombination current is proportional to the recombination rate. 
In particular, Eq.~\eqref{eq:time constants_transient} is a simplification of Eq.~(25) in Ref.~\cite{Boehme03EDMR} including that $\frac{\tau_\text{ap}}{\tau_\text{p}}=0.01\ll 1$, that the $^{31}$P-P$_\text{b0}$ spin-spin coupling is much smaller than the difference of their Larmor frequencies, and that the possibility of spin pair dissociation can be neglected.

In the preceding sections, we have argued that the stretched exponential character observed in most of the measurements is a consequence of a broad distribution of time constants for different spin pairs contributing to the observed signal. Considering in particular $\tau_\text{ap}$, we assume that the recombination involves a tunneling process between the two defects forming the spin pair. Therefore, a distribution of spin pair distances over the ensemble under investigation results in a broad distribution of the recombination time constants $\tau_\text{ap}$, due to the expected exponential dependence of $\tau_\text{ap}$ on the spin pair distance~\cite{Suckert2013}.
As a consequence, the current transient [Eq.~\eqref{eq:time constants_transient}] has to be averaged over a distribution $\rho(r)$ of recombination rates to accurately describe the current transient expected for spin pair ensembles. Assuming a homogeneously $^{31}$P-doped layer of thickness $z_\text{max}$ with a $^{31}$P volume concentration $c$, the area density of dopants $N_\mathrm{D}/A$ is given by
\begin{equation}
\label{eq:Timeconstants_rho}
\begin{split}
\frac{N_\mathrm{D}}{A} & =\int_0^{z_\text{max}}c\cdot\text{d}z \\
& =\int_{r_\text{min}}^{r_\text{max}}\rho(r)\text{d}r,
\end{split}
\end{equation}
where $z$ denotes the distance between the donor and the Si/SiO$_2$ interface, which, for simplicity, we assume to be equal to the $^{31}$P-P$_\text{b0}$ distance. The integration boundaries $r_\text{min}$ and $r_\text{max}$ denote the minimum and maximum recombination rates for the range of $^{31}$P-P$_\text{b0}$ distances considered. We further calculate $\rho(r)$ taking into account an exponential dependence of $r$ on the $^{31}$P-P$_\text{b0}$ distance
\begin{equation}
r(z)=r_0\cdot\text{e}^{-\frac{z}{a}},
\end{equation}
where $r_0$ and $a$ are at first unknown parameters. The change of variables in Eq.~\eqref{eq:Timeconstants_rho} results in
\begin{equation}
\label{eq:Timeconstants_rhoII}
\rho(r)=\frac{c\cdot a}{r}\propto\frac{1}{r}.
\end{equation}

With these considerations, we are now able to calculate the current transient after the detection echo for the inversion recovery experiment as a function of the time interval $T$ (cf.~Fig.~\ref{fig:InvRecDark}). In Sect.~\ref{ss:rateequations}, we have shown that without illumination the time constant of the inversion recovery decay is determined by $r=1/\tau_\text{ap}$, so that
\begin{equation}
\label{eq:Timeconstants_IT}
I(T)\propto \text{e}^{-r\cdot T}.
\end{equation}
Combining Eqs.~\eqref{eq:time constants_transient}, \eqref{eq:Timeconstants_rhoII}, and \eqref{eq:Timeconstants_IT}, the current transient averaged over the distribution of recombination rates is then given by
\begin{equation}
\begin{split}
\left\langle I(t,T)\right\rangle & \propto \int_{r_\text{min}}^{r_\text{max}}\text{e}^{-r\cdot T}\cdot\rho(r)\cdot r\cdot \text{e}^{-r\cdot t} \text{d}r \\
& \propto \int_{r_\text{min}}^{r_\text{max}} \text{e}^{-r\cdot (T+t)}\text{d}r \\
& = -\frac{1}{T+t}\cdot\left(\text{e}^{-r_\text{max}\cdot (T+t)}-\text{e}^{-r_\text{min}\cdot (T+t)}\right).
\end{split}
\end{equation}
This current transient is integrated over the box-car integration interval $\left[t_\text{min},t_\text{max}\right]$ to obtain the charge
\begin{equation}
\label{eq:time constants_DeltaQ}
\begin{split}
\Delta Q(T) &= \int_{t_\text{min}}^{t_\text{max}}\left\langle I(t,T)\right\rangle\text{d}t \\
& = \int_{t_\text{min}}^{t_\text{max}}\frac{1}{T+t}\cdot\left(\text{e}^{-r_\text{min}\cdot (T+t)}-\text{e}^{-r_\text{max}\cdot (T+t)}\right)\text{d}t.
\end{split}
\end{equation}
This integral can be evaluated analytically assuming that $r_\text{min}\cdot(T+t_\text{max})\ll 1$ and $r_\text{max}\cdot(T+t_\text{min})\gg 1$, which means that the range of recombination time constants covered by the distribution $\rho$ is much larger than the timescales of the inversion recovery experiment. With these simplifications, we can evaluate Eq.~\eqref{eq:time constants_DeltaQ} resulting in
\begin{equation}
\label{eq:time constants_DeltaQII}
\begin{split}
\Delta Q(T) &\propto \int_{t_\text{min}}^{t_\text{max}}\frac{1}{T+t}\text{d}t \\
& = \ln\left(\frac{T+t_\text{max}}{T+t_\text{min}}\right).
\end{split}
\end{equation}
\begin{figure}[t!]
\begin{centering}
\includegraphics[width=7cm]{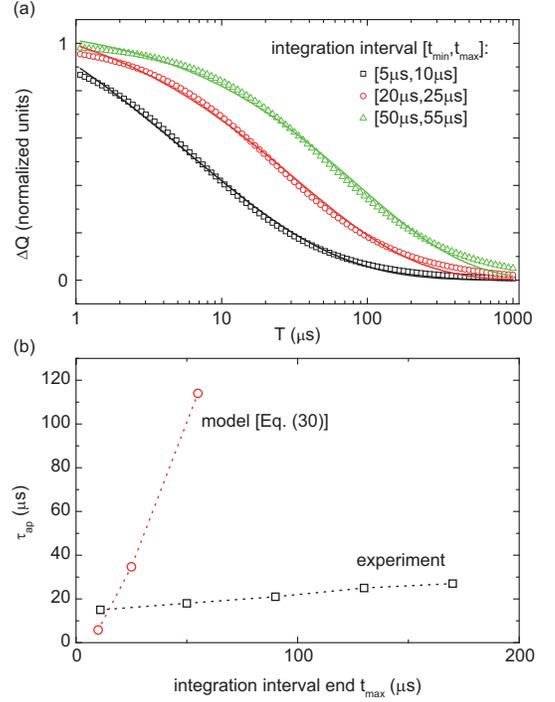} 
\par\end{centering}
\caption{\label{fig:Simu_InvRec}
(a) Plot of Eq.~\eqref{eq:time constants_DeltaQII} as a function of $T$ for different integration intervals (open symbols) and the corresponding stretched exponential fits (solid lines) to determine the characteristic time constant. 
(b) Recombination time of antiparallel spin pairs $\tau_\text{ap}$ for different box-car integration interval $\left[t_\text{max}-5~\mu\text{s},t_\text{max}\right]$ measured by an inversion recovery with pulsed illumination (black squares). For comparison, the time constants of the stretched exponential fits in (a) are shown as well (red circles). The dashed lines are a guide to the eye.
}
\end{figure}
Equation~\eqref{eq:time constants_DeltaQII} describes a decay which closely resembles a stretched exponential, as shown in Fig.~\ref{fig:Simu_InvRec}(a), where Eq.~\eqref{eq:time constants_DeltaQII} is plotted for different integration intervals $\left[t_\text{min},t_\text{max}\right]$ (open symbols). The solid lines are stretched exponential fits with time constants of 5.8~$\mu$s, 34.8~$\mu$s, and 114~$\mu$s and exponents of 0.41, 0.52, and 0.58 for $\left[t_\text{min},t_\text{max}\right]=$[5~$\mu$s,10~$\mu$s], [20~$\mu$s,25~$\mu$s], and [50~$\mu$s,55~$\mu$s], respectively. From Eq.~\eqref{eq:time constants_DeltaQII} and Fig.~\ref{fig:Simu_InvRec} it is clear that if the assumptions hold for which Eq.~\eqref{eq:time constants_DeltaQII} has been derived, the time constant measured in an inversion recovery experiment is completely determined by the box-car integration interval, and therefore independent of the spin pair properties.
Indeed, a dependence of the inversion recovery time constant on the integration interval is observed for the $^{31}$P-P$_\text{b0}$ spin system for an inversion recovery under pulsed illumination. Figure~\ref{fig:Simu_InvRec}(b) shows $\tau_\text{ap}$ for different box-car integration interval $\left[t_\text{max}-5~\mu\text{s},t_\text{max}\right]$ obtained from a fit as described in Sect.~\ref{ss:rs} with $\tau_\text{p}$=1.2~ms fixed. The observed value of $\tau_\text{ap}$ increases with increasing $t_\text{max}$ from 15~$\mu$s to 27~$\mu$s for $t_\text{max}$=11~$\mu$s to $t_\text{max}$=170~$\mu$s, although with a much weaker dependence than predicted by Eq.~\eqref{eq:time constants_DeltaQII}.

The considerations leading to Eq.~\eqref{eq:time constants_DeltaQII} certainly involve several simplifications. Most importantly, the time dependence of the current transient is more complicated than assumed in Eq.~\eqref{eq:time constants_transient}, including also the electron and hole capture time constants as discussed in the previous section. In particular, the decay of the current transient for short $\tau_\mathrm{ap}$ is determined by $\tau_\text{ec}$ and $\tau_\text{hc}$ rather than $\tau_\mathrm{ap}$. The data shown in Fig.~\ref{fig:transients} was recorded for an illumination intensity of 20~mW/cm$^2$, corresponding to $\tau_\text{ec}\approx$50~$\mu$s\textgreater $\tau_\text{ap}$=4~$\mu$s. We therefore expect that $\Delta Q$ only weakly depends on $\tau_\text{ap}$, explaining the weak dependence of $\tau_\text{ap}$ on $t_\text{max}$ observed in the experiment.
However, in this case, the measurement of the electron capture time constant $\tau_\text{ec}$ is expected to show a stronger dependence on the integration interval if the values of $\tau_\text{ec}$ vary sufficiently over the spin pair ensemble.

In conclusion, we have demonstrated that the choice of the boxcar integration interval can have a pronounced influence on the recombination time constants measured by pulsed EDMR. In particular, for a simplified model of the photocurent
transient, we have shown that the time constant
observed in a inversion recovery experiment is completely
determined by the boxcar integration interval. In the
experimental data however, the dependence is much
weaker than predicted by the model indicating that these
time constants depend on the properties of the spin pair.
But care has to be taken when interpreting these
results.

\section{Summary}
We developed pulsed electrically detected magnetic resonance measurements combined with pulsed optical excitation to characterize several time constants involved in the spin-dependent recombination process via $^{31}$P-P$_\text{b0}$ spin pairs at the Si/SiO$_2$ interface. In the particular samples studied, we determined the recombination times of parallel and antiparallel spin pairs and obtain values of $\tau_\text{ap}$=15.5(8)~$\mu$s and $\tau_\text{p}$=1.2(1)~ms, respectively. The recombination of parallel spin pairs is attributed to a spontaneous spin flip of the P$_\text{b0}$. We also measured the generation time of new spin pairs, which we find to depend linearly on the illumination intensity. Using pulsed ENDOR, we identified this generation time constant with the capture time constant for electrons from the conduction band by $^{31}$P$^+$ donors. Based on these time constants, we devised a set of rate equations to calculate the current transient after a resonant microwave pulse and compare it with experimental data, which allowed to also estimate the hole capture rate. We further demonstrated experimentally and theoretically that the choice of the boxcar integration interval influences the recombination time constants measured by pulsed EDMR. The reason for this appears to be the broad distribution of time constants caused by the variation of $^{31}$P-P$_\text{b0}$ distances over the spin pair ensemble, which is also reflected in the stretched exponential character of the observed decays.  

The experimental values obtained are, as pointed out, specific to the samples studied. It would, therefore, be important to apply the techniques developed here to a systematic study of the $^{31}$P-P$_\text{b0}$ recombination process by varying the concentration of interface states, e.g., dry or wet oxides and H-terminated or organically functionalized surfaces. It would be interesting to study spin pairs with different distributions of time constants, e.g., by varying the thickness of the doped epilayer and ultimately by confining the dopants to a monoatomic layer at a certain distance from the Si/SiO$_2$ interface~\cite{Schofield03STMdonor,McKibbin2009}. Furthermore, the presented techniques can be applied to other spin pairs involving, e.g., the P$_\mathrm{b1}$ center at the Si/SiO$_2$ interface which is controversely discussed in the literature~\cite{Stesmans98dbDensity,Mishima99}, defects in thin-film silicon solar cells~\cite{Schnegg2012} or spin-dependent transport in organic semiconductors~\cite{Dyakonov1996,Mccamey08,Behrends2010}. 

\vspace{1cm}

\begin{acknowledgments}
The work was financially supported by DFG (Grant No. SFB 631, C3 and Grant No. SPP1601, Br 1585/8-1) with additional support by BMBF via EPR-Solar. 
\end{acknowledgments}

\appendix
\section{Fill Pulse Length \label{Appendix_FPL}}

In Sect.~\ref{ss:generation}, we measured the spin echo amplitude as a function of the length $T_\mathrm{LED}$ of the illumination fill pulse to determine the generation rate of new spin pairs $1/\tau_\text{g}$ (Fig.~\ref{fig:FPL}).
In this appendix, we derive a formula for the phase-cycled spin echo amplitude as a function of $T_\mathrm{LED}$ and show that for $1/\tau_\text{ap}\gg 1/\tau_\text{g},1/\tau_\text{p}$, the time dependence is determined only by the time constant $1/\lambda_1$ given in Eq.~\eqref{eq:time constants_eigenvalues_simple}.

We assume that before the illumination fill pulse all spin pairs have recombined, so that the state vector $\rho(0)$ is given by
\begin{equation}
\rho(0)=
\begin{pmatrix}
	0\\0\\1
\end{pmatrix}.
\label{eq:Appendix_FPL_rho0}
\end{equation}
After an illumination pulse of length $T$ the populations have evolved to
\begin{equation}
\rho(T)=e^{\tilde{R}\cdot T} 
\begin{pmatrix}
	0\\0\\1
\end{pmatrix}.
\label{eq:Appendix_FPL_rhoT}
\end{equation}
By carrying out the matrix exponential, we obtain an expression for the spin echo amplitude $\Delta Q$ given by 
\begin{widetext}
\begin{eqnarray}
\Delta Q\propto n_\mathrm{ap}-n_\mathrm{p} =&
\mbox{\large$ -\frac{\frac{1}{\tau_\text{g}}(\frac{1}{\tau_\text{ap}}-\frac{1}{\tau_\text{p}})}{\frac{2}{\tau_\text{ap}\tau_\text{p}}+\frac{1}{\tau_\text{g}}(\frac{1}{\tau_\text{ap}}+\frac{1}{\tau_\text{p}})}$}\nonumber\\
&
\mbox{\large$
-\frac{\frac{1}{\tau_\text{g}}(\frac{1}{\tau_\text{ap}}-\frac{1}{\tau_\text{p}})}{(\frac{1}{\tau_\text{g}})^2+(\frac{1}{\tau_\text{ap}}-\frac{1}{\tau_\text{p}})^2-(\frac{1}{\tau_\text{g}}+\frac{1}{\tau_\text{ap}}+\frac{1}{\tau_\text{p}})\sqrt{(\frac{1}{\tau_\text{g}})^2+(\frac{1}{\tau_\text{ap}}-\frac{1}{\tau_\text{p}})^2}}e^{\lambda_1 t}$}\nonumber\\
&
\mbox{\large$
-\frac{\frac{1}{\tau_\text{g}}(\frac{1}{\tau_\text{ap}}-\frac{1}{\tau_\text{p}})}{(\frac{1}{\tau_\text{g}})^2+(\frac{1}{\tau_\text{ap}}-\frac{1}{\tau_\text{p}})^2+(\frac{1}{\tau_\text{g}}+\frac{1}{\tau_\text{ap}}+\frac{1}{\tau_\text{p}})\sqrt{(\frac{1}{\tau_\text{g}})^2+(\frac{1}{\tau_\text{ap}}-\frac{1}{\tau_\text{p}})^2}}e^{\lambda_2 t},$}
\label{eq:Appendix_FPL_DeltaRho}
\end{eqnarray}

\end{widetext}
where $n_\mathrm{p}$ and $n_\mathrm{ap}$ are given by the first and second component of $\rho$, respectively (c.f. Eq.~\eqref{eq:rho}).
For $\tau_\text{ap}\ll \tau_\text{g},\tau_\text{p}$, Eq.~\eqref{eq:Appendix_FPL_DeltaRho} simplifies to
\begin{equation}
n_\mathrm{ap}-n_\mathrm{p}=-\frac{1}{1+2\frac{\tau_\text{g}}{\tau_\text{p}}}+\frac{1}{1+2\frac{\tau_\text{g}}{\tau_\text{p}}}e^{\lambda_1 t}-\frac{\tau_\text{ap}}{2\tau_\text{g}}e^{\lambda_2 t}.
\label{eq:Appendix_FPL_DeltaRho_Simple}
\end{equation}
The third term on the right hand side of Eq.~\eqref{eq:Appendix_FPL_DeltaRho_Simple} is suppressed compared with the second term by a factor of
\begin{equation}
\frac{\tau_\text{ap}}{2\tau_\text{g}}(1+2\frac{\tau_\text{g}}{\tau_\text{p}})<0.1
\label{eq:Appendix_thirdTerm}
\end{equation}
for the experiments in this work (cf. Fig.~\ref{fig:InvRecDark}, Fig.~\ref{fig:g_summary}, and Fig.~\ref{fig:DPS}), so that we can neglect this term in Eq.~\eqref{eq:Appendix_FPL_DeltaRho_Simple}. This justifies to fit the data in Fig.~\ref{fig:FPL} with a single exponential with a time constant $\lambda_1$.

\end{document}